\documentclass[fleqn,usenatbib]{mnras}

\usepackage{newtxtext,newtxmath}

\usepackage[T1]{fontenc}

\DeclareRobustCommand{\VAN}[3]{#2}
\let\VANthebibliography\thebibliography
\def\thebibliography{\DeclareRobustCommand{\VAN}[3]{##3}\VANthebibliography}

\usepackage{graphicx}	
\usepackage{amsmath}	
\usepackage{caption}
\usepackage{float}

\title[Relativistic orbital deformation from improved timing of PSR~J1757$-$1854]{Detection of relativistic orbital deformation from improved timing of PSR~J1757$-$1854}
\author[Singha et al.]{\parbox{\textwidth}{
Jaikhomba Singha,$^{1}$\thanks{E-mail: mjaikhomba@gmail.com }
Vivek Venkatraman Krishnan$^{2}$,
Marisa Geyer$^{1}$,
Victoria Blackmon$^{3,4}$,
Paulo Freire$^{2}$,
Norbert Wex$^{2}$,
Maura McLaughlin$^{3,4}$,
Michael Kramer$^{2,5}$, 
Amanda Weltman$^1$,
David Champion$^{2}$,
Matthew Bailes$^{6}$,
Sarah Buchner$^{7}$,
Fernando Camilo$^{7}$,
Andrea Possenti$^{8}$,
Maciej Serylak$^{9}$.}
\\\\
$^{1}$High Energy Physics, Cosmology \& Astrophysics Theory Group (HEPCAT), Department of Applied Mathematics, University of Cape Town, South Africa\\
$^{2}$Max-Planck-Institut f{\"u}r Radioastronomie, Auf dem H{\"u}gel 69, D-53121 Bonn, Germany.\\
$^{3}$Department of Physics and Astronomy, West Virginia University, PO Box 6315, Morgantown, WV 26506, USA\\
$^{4}$Center for Gravitational Waves and Cosmology, West Virginia University, Chestnut Ridge Research Building, Morgantown, WV 26505, USA \\
$^{5}$Jodrell Bank Center for Astrophysics, University of Manchester, M13 9PL, UK\\
$^{6}$Centre for Astrophysics and Supercomputing, Swinburne University of Technology, ARC Centre of Excellence for Gravitational Wave Discovery (OzGrav)\\
$^{7}$South African Radio Astronomy Observatory, Liesbeek House, River Park, Cape Town 7705, South Africa\\
$^{8}$INAF – Osservatorio Astronomico di Cagliari, via della Scienza 5, 09047 Selargius (CA), Italy\\
$^{9}$SKA Observatory, Jodrell Bank, Lower Withington, Macclesfield, Cheshire, SK11 9FT, UK.}

\date{Accepted XXX. Received YYY; in original form ZZZ}

\pubyear{\the\year{}}

\begin{document}
\label{firstpage}
\pagerange{\pageref{firstpage}--\pageref{lastpage}}
\maketitle

\begin{abstract}
PSR~J1757$-$1854, a 21.5\,ms pulsar is a member of a highly relativistic double neutron star (DNS) system with a tight (4.4 hr), eccentric ($e = 0.61$) orbit. With extremely large gravitational wave luminosity and one of the fastest orbital decay rates of any known DNS system, it is ideal for testing general relativity (GR) in the strong-field regime. Here we present results from a high-precision timing campaign combining archival data from the Murriyang telescope and Green Bank Telescope (GBT) with new observations from the MeerKAT radio telescope and additional observations from the GBT. The extended baseline and superior sensitivity of MeerKAT have yielded substantial improvements to previously measured post-Keplerian parameters by a factor of around $\sim2$ or more. We report the first detection of relativistic angular deformation, $\delta_\theta$ in this system, making PSR~J1757$-$1854 only the third DNS system for which $\delta_\theta$ has been measured, achieved here in just 9 yrs compared to the decades of timing required for both the double pulsar and the Hulse-Taylor binary. We demonstrate how $\delta_\theta$ can be used to constrain the spin-orbit geometry of the system, ruling out two of the four geometric solutions previously identified, while remaining consistent with GR. We also evaluate higher-order contributions to the periastron advance $\dot{\omega}$, including the second post-Newtonian correction and the Lense-Thirring term, and show that these have a measurable systematic effect on the inferred total system mass. The observed orbital period derivative, $\dot{P}_\mathrm{b}$ remains consistent with the GR prediction for gravitational-wave damping across a wide range of plausible distances.
\end{abstract}




\section{Introduction}

Pulsars are rapidly rotating neutron stars whose remarkable rotational stability renders them among the most precise natural clocks in the Universe. Since their discovery, they have proven to be powerful tools for probing fundamental physics across a wide range of regimes \citep{lorimer_kramer2005}.

\subsection{Binary pulsars and general relativity}

Among the diverse population of pulsars, those residing in compact binary systems with another neutron star, known as double neutron star (DNS) systems, occupy a uniquely privileged position. The gravitational fields in such systems are sufficiently strong and the orbital velocities sufficiently high that the orbital dynamics deviate measurably from the predictions of Newtonian gravity, providing a direct observational window onto the strong-field regime of gravitational physics. This allows us to probe gravity theories, especially general relativity (GR), in this regime. Precise timing of the DNS system, PSR~B1913+16, provided the first observational evidence for the emission of gravitational waves, through the measured decay of the orbital period in precise agreement with the prediction of GR \citep{Taylor_etal1979, taylor1982new}. Since then, a growing number of DNS systems have been discovered and timed with increasing precision, including the ``double pulsar" system, which provides a 25-fold improvement in the precision of the radiative properties of gravity and of the propagation of light \citep{kramer2021}. Such tests are now being done with even more extreme binaries \citep{cameron2023new,Meng_2025}. This and other systems are providing new and complementary tests of GR and alternative theories of gravity \citep[for a review, see ][]{FreireWex2024}.

The formation and evolution of DNS systems is a complex astrophysical process. Following the initial supernova explosion that produces the first-born neutron star, the system enters a phase in which the neutron star accretes matter from its evolving companion, spinning itself up to millisecond rotation periods in a process known as recycling \citep{Radhakrishnan_Srinivasan1982}. The idea of recycling for PSR B1913+16 was first suggested by \citet{Smarr_Blandford1976}, and confirmed by the work of \citet{Srinivasan_vandenHeuvel1982}, demonstrating for the first time that the companion of PSR B1913+16 must be a neutron star younger than the observed pulsar. The model for the formation of PSR B1913+16 and other close DNS, as a post-common-envelop product of a High-Mass X-ray Binary, was put forward by \citet{Flannery_vandenHeuvel1975} and \citet{vandenHeuvel1976}. The model for the evolution of Low-Mass X-ray Binary systems was proposed by \citep{Alpar_etal1982}. When the companion exhausts its nuclear fuel and undergoes its own supernova explosion, a DNS binary is formed, typically characterised by a mildly recycled pulsar and a younger, non-recycled companion neutron star. The second supernova imparts a natal kick to the newly formed neutron star, which can significantly misalign the pulsar spin axis with respect to the orbital angular momentum vector \citep{tauris2017dns, Vigna_etal2018, Andrews_Mandel2019}. The resulting systems are among the most extreme gravitational environments accessible to observational study, and the properties of the DNS orbit encode information about the supernova mechanism and the prior evolutionary history of the binary \citep{Tauris_etal2015, Vigna_etal2018}.

The technique of pulsar timing provides the means to exploit these extreme environments for fundamental physics. In this approach, a timing model describing the rotational, astrometric, and orbital properties of the pulsar is fitted to the pulse times of arrival (ToAs), and the residuals between the observed and model-predicted arrival times are minimised \citep{lorimer_kramer2005, HobbsTempo22006}. The orbital dynamics of a DNS system is described to leading order by the five Keplerian parameters: the orbital period ($P_\mathrm{b}$), the projected semi-major axis of the pulsar orbit ($x$), the orbital eccentricity ($e$), the longitude of the periastron ($\omega$) and the epoch of periastron passage ($T_0$). These parameters can be determined with extraordinary precision through long-term monitoring campaigns. The precision of the derived parameters improves with different powers of the length of the timing baseline $t$ \citep{damour1986general}, making sustained, high-cadence timing observations particularly powerful.

Beyond the classical Keplerian description, pulsar timing enables measurements of relativistic deviations from Newtonian orbital motion, parametrised through a set of phenomenological, theory-independent parameters known as the post-Keplerian (PK) parameters \citep{damour1986general,damour1992strong}. Within any given theory of gravity, they are uniquely determined by the component masses $m_\mathrm{p}$ and $m_\mathrm{c}$ and the Keplerian orbital elements, but their functional forms differ between gravity theories \citep{damour1986general, damour1992strong, FreireWex2024}. Some of the most commonly measured PK parameters are: (i) the relativistic advance of periastron ($\dot{\omega}$), arising from the first post-Newtonian corrections to the equations of motion of a binary system; (ii) the amplitude of the Einstein delay ($\gamma$), which encodes the combined effects of the second-order Doppler shift and gravitational redshift as the pulsar moves through the gravitational potential of its companion; (iii) the orbital period derivative ($\dot{P}_\mathrm{b}$), which encodes the loss of energy from the system due to the emission of gravitational waves; (iv) the Shapiro delay range parameter ($r$) and (v) shape parameter ($s$), which together describe the additional propagation delay experienced by the pulsar signal as it passes through the gravitational well of the companion. Other PK parameters have never, or very rarely been measured, such as (vi) the relativistic radial deformation parameter ($\delta_r$); and (vii) the relativistic angular deformation parameter ($\delta_\theta$), these arise from purely periodic relativistic corrections to the Keplerian motion \citep{damour1986general, damour1992strong}. The measurement of any two independent PK parameters allows the individual component masses to be determined uniquely within the framework of a given theory of gravity. When three or more PK parameters are available, each additional PK parameter constitutes an independent, self-consistency test of the gravitational theory \citep{Taylor_NobelLecture1993, FreireWex2024}. To date, GR has passed every such test with remarkable precision.

The most compact DNS systems provide access not only to the suite of PK parameters described above, but also to additional relativistic effects that remain inaccessible in less extreme binaries. Among the most important of these are the geodetic precession (GP) and Lense-Thirring (LT) effect; both arise from the relativistic coupling between the spin angular momentum of the pulsar and the orbital angular momentum of the system \citep{Barker_Connel1975}.

Geodetic precession is the effect of relativistic spin-orbit coupling on the spin of the pulsar. It causes the pulsar spin axis to precess around the total angular momentum vector of the system at a rate determined by the component masses and orbital parameters. This precession manifests observationally as a secular evolution of the pulse profile morphology, as the changing orientation of the spin axis alters the line-of-sight cut through the pulsar emission beam, allowing unique constraints on emission beam structure \citep{kramer1998geometry, kramer2003geodetic, breton2008precession, desvignes2012geometry}.

The LT effect is the effect of relativistic spin-orbit coupling on the orbit. It contributes measurably to several PK parameters, most notably the periastron advance $\dot{\omega}$ and a secular change in the projected semi-major axis $\dot{x}$ \citep{Barker_Connel1975, Damour_Schafer1988, norbert1995secondPN, FreireWex2024}. Precise measurements of the LT contribution are of particular astrophysical significance for two reasons. First, they provide an independent test of GR in the strong-field regime by comparing the measured spin-orbit coupling with the theoretical prediction \citep{damour1992strong, FreireWex2024}. Second, and perhaps more importantly, assuming GR is correct, they enable direct observational estimates of the neutron star moment of inertia $I$, which is a sensitive discriminant of the equation of state (EOS) of ultra-dense nuclear matter (e.g. \citealt{Lattimer_Prakash2001}). To date, the LT contribution to the PK parameters in a DNS has been seen only in the double pulsar system PSR~J0737$-$3039A/B \citep{kramer2021}, where it has provided the first observational constraint on the moment of inertia of a neutron star through such measurements.

\subsection{PSR~J1757\texorpdfstring{$-$}{-}1854}
In this work, we present results from a high-precision timing campaign for the highly relativistic DNS system PSR~J1757$-$1854, one of the most extreme binary pulsar systems currently known. PSR~J1757$-$1854 is a mildly recycled pulsar with a spin period of 21.5\,ms \citep{cameron2018most} discovered in the High Time Resolution Universe South (HTRU-S) Galactic plane pulsar survey conducted with the 64\,m Murriyang (previously Parkes) radio telescope \citep{kjvs+10}. The system has an orbital period of just 4.4\,hr and an orbital eccentricity of $e=0.61$ \citep{cameron2023new}, making it one of the most compact and eccentric DNS systems in the Milky Way. The compactness and eccentricity of this system give rise to an extraordinary set of physical conditions at periastron passage. The acceleration reaches $\sim 680$\,m\,s$^{-2}$, the relative orbital velocity reaches $\sim 1060$\,km\,s$^{-1}$, and the binary separation at periastron is only $0.749\,\mathrm{R}_\odot$. These values represent the most extreme known among all DNS systems. As a direct consequence, PSR~J1757$-$1854 has one of the highest known gravitational wave luminosities and also one of the fastest orbital decay rates, $|\dot{P}_b|$ due to gravitational wave damping, with a merger timescale of approximately 76\,Myr \citep{cameron2018most}. This places it among the DNS systems that will merge within a Hubble time and contribute to estimating the population of binary NS mergers.

\begin{table*}
\caption{Details of the data used in this work taken from \protect\cite{cameron2023new}, in addition to new data obtained using MeerKAT and GBT (given in boldfonts). The table lists the various receivers, backends and the corresponding data span, and the central frequency ($f_\text{c}$) and bandwidth ($\Delta f$) of each receiver/backend combination. The median ToA uncertainties have been quoted in the last column. Note that the GBT - New data uses the same GBT backend as used in one of the datasets in \protect\cite{cameron2023new}, but it is mentioned here separately to indicate that this was added to the previous data.}\label{tab:observations}
\begin{center}
\begin{tabular}{lllllll}
\hline
\hline
Telescope & Receiver & Backend & Span & $f_\text{c}$ & $\Delta f$ & Median $\sigma_r$\\
 & & & (MJD) & (MHz) & (MHz) & ($\mu$s) \\
\hline
\textit{Murriyang (previously Parkes)} & MB20 & BPSR & 57405--57406 & 1382 & 400 & 93.9 \\
 & & DFB4 & 57734--58674 & 1369 & 256  & 98.7 \\
 & & CASPSR & 57734--58674  & 1382 & 400  & 63.6 \\
 & H-OH & DFB4 & 57450--57676  & 1369 & 256 & 62.3 \\
 & & CASPSR & 57596--57651 & 1382 & 400 & 58.1 \\
 & UWL & DFB4 & 58438--58890  & 1369 & 256  & 61.2 \\
 & & CASPSR & 58793--58890  & 1382 & 400 & 38.5\\
\hline
\textit{GBT} & PF1-800 & GUPPI & 57620--57621  & 820 & 200 & 140.8  \\ 
& L-Band & GUPPI & 57795--58863  & 1499 & 800 & 37.7 \\
 & & VEGAS & 58165--59594 & 1501 & 800  & 33.5\\
 & S-Band & GUPPI & 57627--58862  & 1999 & 800 & 28.8 \\
 & & VEGAS & 58165--59593  & 2001 & 800  & 30.4\\
\hline
\textit{\textbf{MeerKAT}} & \textbf{L-Band} & \textbf{PTUSE} & \textbf{58550--60293}  & \textbf{1284} & \textbf{856} & \textbf{52.2}  \\
 & \textbf{S1-Band} & \textbf{PTUSE} & \textbf{60077--60386}  & \textbf{2406} & \textbf{875} & \textbf{25.9} \\
\hline
\textit{\textbf{GBT - New data}} & \textbf{L-Band} & \textbf{VEGAS} & \textbf{60355--60757}  & \textbf{1501} & \textbf{800} & \textbf{40.7} \\
\hline
\hline
\end{tabular}    
\end{center}
\end{table*}
Since its discovery, PSR~J1757$-$1854 has been the subject of sustained timing campaigns with the Murriyang (previously Parkes) telescope and the Green Bank Telescope (GBT), see Table~\ref{tab:observations}. Six years of observations with these facilities yielded measurements of five PK parameters \citep{cameron2023new}: $\dot{\omega}$, $\gamma$, $\dot{P}_\mathrm{b}$, and the two 
orthometric Shapiro delay parameters $h_3$ and $\varsigma$. The orthometric parametrisation \citep{freire2010orthometric} replaces the standard Shapiro delay range and shape parameters $r$ and $s$ with a combination of parameters that are less correlated and more robustly measurable for almost all systems except those with the highest orbital inclinations, such as the double pulsar \citep{kramer2021}. 
All PK parameters measured by \cite{cameron2023new} are consistent with GR except the intrinsic $\dot{P}_\mathrm{b}$, which has a large systematic uncertainty. This is a result of the unknown distance to the system, thereby providing large uncertainties in the extrinsic contributions to the observed $\dot{P}_\mathrm{b}$ \citep{cameron2023new}.

In the present work, we add data from MeerKAT telescope and GBT to the 6 yr of existing data to increase the timing baseline to 9 yr. This longer baseline and highly sensitive data from MeerKAT resulted in improved measurements of the PK parameters which now include the relativistic orbital deformation. PSR~J1757$-$1854 is expected to be a prime candidate for the measurement of the LT effect \citep{Bagchi2018, Hu_Freire2024}: the detection of the effects of GP in this system \citep{cameron2023new} indicate a significant misalignment between the spin of the pulsar and the orbital angular momentum of the system. This is a necessary condition for the detection\footnote{These are the observed orbital quantities and not the actual ones since they are the results of unmodelled aberration effects.} of  the derivative of the projected semi-major axis, $\dot x$ and that of the eccentricity, $\dot e$ and the measurement of moment of inertia. 

The remainder of this paper is organised as follows. Section~\ref{Sec:Obs-datareduction} describes the observing setup, data acquisition, and data reduction procedures for the MeerKAT and GBT datasets. Section~\ref{Sec:TimingAnalysis} presents the timing analysis methodology, including the noise modelling framework. The results of the timing analysis, including updated parameter measurements, the detection of $\delta_\theta$, the analysis of $\dot{P}_\mathrm{b}$ contributions, and the assessment of higher-order corrections to $\dot{\omega}$, are presented and discussed in Section~\ref{Sec:ResultsDiscussion}. Section~\ref{Sec:ConclusionFuture} summarises our conclusions and future scope of this system.

\begin{figure*}
 \centering
  \includegraphics[scale=.55]{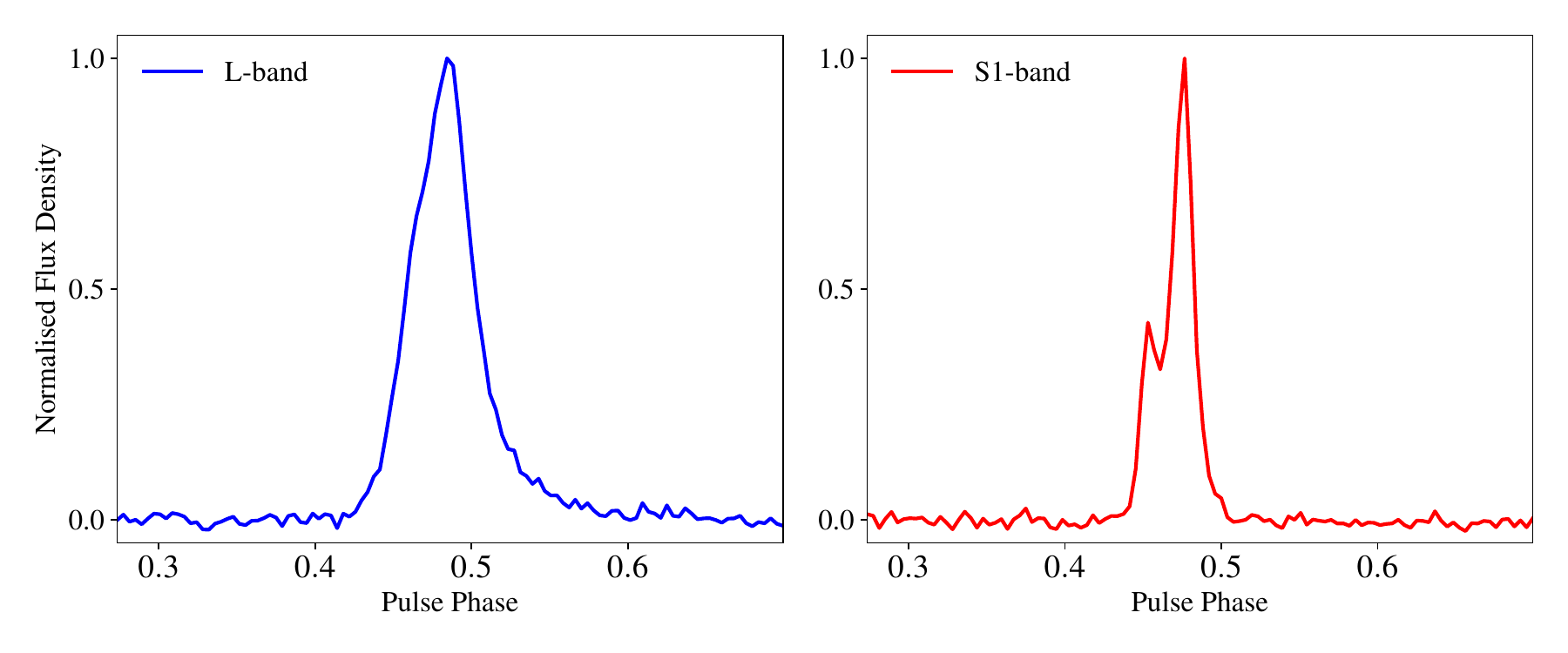}
  \caption{Representative pulse profiles of PSR~J1757$-$1854 observed with the MeerKAT telescope at L-band (left panel, 1284\,MHz) and S1-band (right panel, 2406\,MHz), each obtained from a single observing epoch. Both profiles are normalised to a peak flux density of unity. The L-band profile displays a pronounced asymmetric scattering tail, a characteristic signature of multi-path scattering in the turbulent interstellar medium along the line of sight. This broadening is substantially reduced at S1-band frequencies, where the intrinsic pulse shape is better preserved. The narrower and more symmetric S1-band profile highlights the clear advantage of higher frequency observations for achieving the high timing precision for this heavily scattered, high DM pulsar.}
  \label{profile}
\end{figure*}


\section{Observations and Data Reduction}
\label{Sec:Obs-datareduction}
The observations of PSR~J1757$-$1854 were carried out across multiple frequency bands and observing epochs, the details of which are summarised in Table~\ref{tab:observations}. Below we describe the observing setup and data reduction procedures for each facility in turn.
\subsection{Observations with MeerKAT} 
The MeerKAT radio telescope is a highly sensitive interferometric array consisting of 64 offset Gregorian dishes, each with an effective diameter of 13.9\,m, situated in the Karoo region of the Northern Cape province of South Africa \citep{Jonas2016MeerKAT}. The telescope is operated by the South African Radio Astronomy Observatory (SARAO) and serves as a mid-frequency precursor instrument to the Square Kilometre Array (SKA). The array has a maximum baseline of approximately 8\,km, and in tied-array beamforming mode, which is the configuration used for pulsar timing, the signals from all dishes are coherently summed to produce a single high-sensitivity beam directed at the target source. Owing to its large total collecting area, low receiver noise temperatures, and the exceptionally low radio frequency interference (RFI) environment of the Karoo site, MeerKAT is among the most sensitive radio telescopes in the world for pulsar timing \citep{BailesMeerKAT2020}. These properties make it particularly well suited for high-precision timing of faint or heavily dispersed pulsars such as PSR~J1757$-$1854, which has a dispersion measure (DM) of $\sim 378$\,pc\,cm$^{-3}$ and lies at low Galactic latitude ($b \sim 2.9^\circ$), where propagation effects are significant and can introduce frequency-dependent structure to the pulse profile and systematic offsets into the measured ToAs. 

PSR~J1757$-$1854 was observed with meerKAT as part of the Relativistic Binary programme (RelBin: \citealt{Kramer_etal2021RelBin}), a dedicated sub-theme of the MeerTime Large Survey Project \citep{BailesMeerKAT2020} on the MeerKAT radio telescope. The primary scientific goal of the RelBin programme is to perform high-precision pulsar timing of the most relativistic binary systems visible from the Southern Hemisphere, with the overarching aim of measuring post-Keplerian parameters, conducting stringent tests of GR and obtaining NS masses. Due to its extreme relativistic properties, PSR~J1757$-$1854 is among the highest-priority targets of this programme and has been observed with high cadence for around 5 yrs (for details, see Table~\ref{tab:observations}).

The system was observed in two frequency bands resulting in improved timing precision and sensitivity to frequency-dependent effects such as DM variations and interstellar scattering. The L-band, centred at 1284\,MHz with a total bandwidth of 856\,MHz (spanning 856--1712\,MHz), provides high sensitivity and is well suited for measuring the dispersive and scattering properties of the interstellar medium along the line of sight. The MeerKAT S-band receivers have five bands, each with a bandwidth of 875 MHz. Here we use S1 band, centred at 2406\,MHz and thus spanning 1968--2843\,MHz, provides higher timing precision due to the reduced influence of interstellar dispersion and scattering at higher frequencies, and allows for precise chromaticity measurements.

The data were recorded using the Pulsar Timing User Supplied Equipment (PTUSE) backend \citep{BailesMeerKAT2020}, which is the primary beamforming and pulsar timing backend of MeerKAT. PTUSE receives the coherently summed tied-array beam from the MeerKAT correlator and produces coherently dedispersed, full-Stokes, folded pulse profiles in real time. Coherent dedispersion removes the dispersive smearing of the pulse across the observing band, ensuring that the intrinsic pulse shape is preserved and that the achievable timing precision is not limited by intra-channel dispersive broadening. This is particularly important for PSR~J1757$-$1854 given its high DM, where the dispersive smearing across a single frequency channel would otherwise be significant. Each observation was recorded using dual polarisation feeds, such that we are able to produce the full set of Stokes parameters, enabling polarisation calibration and the study of the polarimetric properties of the pulse profile if required.

The raw folded data products from PTUSE were transferred to the OzStar high-performance computing facility at Swinburne University of Technology, Melbourne, Australia, where the initial data reduction were carried out. Initial data reduction was performed using the  \textsc{meerpipe} pipeline \citep{BailesMeerKAT2020}, which implements a standardised and automated workflow for MeerKAT pulsar timing data. A detailed description of the MeerKAT observing system, the PTUSE backend, and the \textsc{meerpipe} pipeline can be found in \cite{BailesMeerKAT2020}. The MeerKAT observations included in this analysis span a timing baseline of approximately five years (MJD 58550--60386), with a total integration time of approximately 4.5\,hr per epoch. Following the initial pipeline reduction, the \textsc{psrchive} \citep{HotanPSRCHIVE2004} utility \textit{pam} was used to further scrunch the data in time to sub-integration lengths of $\sim 6$\,min, in frequency to 8 sub-bands and in phase bins to 256 bins per epoch providing the time and frequency resolution required for precise ToA extraction.  

Representative pulse profiles obtained from a single observing epoch at both frequency bands are shown in Fig.~\ref{profile}. The effect of interstellar scattering is clearly visible in the L-band profile, which exhibits a pronounced scattering tail that significantly broadens the pulse compared to the S1-band profile. This frequency-dependent broadening is consistent with the high DM and low Galactic latitude of the system \citep{NE2001, Bhat_etal2004, NE2025} and underscores the value of S1-band observations for achieving the highest timing precision for this pulsar.  The typical signal-to-noise ratio (S/N) achieved in a single observation (one orbit) averaged over the full frequency band is approximately 166 in the L-band and 137 in the S1-band, reflecting the excellent sensitivity of MeerKAT for this target at both frequency bands. These high S/N values 
make the MeerKAT data (particularly the S1-band data) the highest-quality component of the combined  dataset used in this work.
\begin{figure*}
  \centering
  \includegraphics[scale=.75, trim={0 0 0 0},clip]{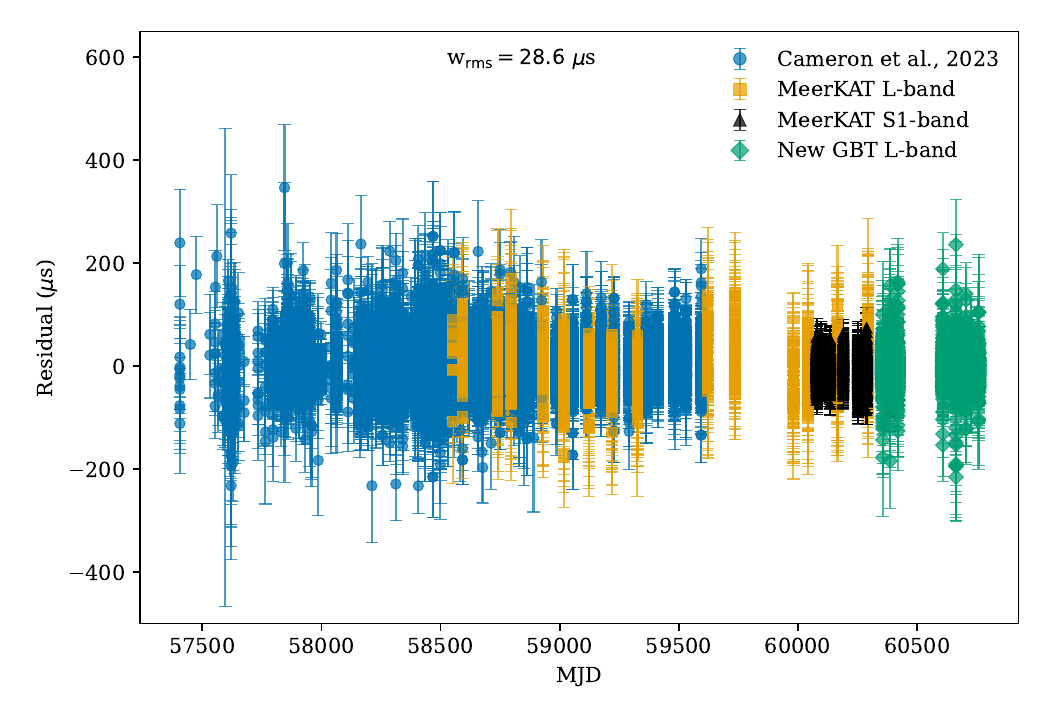}
  \caption{Timing residuals of PSR~J1757$-$1854 as a function of Modified Julian Date (MJD), obtained from the combined dataset spanning 9 yrs. Blue circles represent the archival Murriyang (previosuly Parkes) and GBT data from \protect\cite{cameron2023new}, orange squares and =black triangles are the new MeerKAT L-band and S1-band observations respectively, and bluish-green diamonds are the new data using GBT. The vertical extent of the residuals in the earlier \protect\cite{cameron2023new} dataset reflects the larger ToA uncertainties of those observations. The visibly reduced scatter and smaller error bars in the MeerKAT S1-band data compared to the L-band data are consistent with the frequency-dependent reduction in scattering at higher frequencies. The data presented spans around 9 years, resulting in a weighted rms timing residual, w$_{\rm rms}$ = 28.6$~\mu$s.} 
  \label{fig:timing-residuals}
\end{figure*}
\subsection{Observations with the Green Bank Telescope}
New observations of PSR~J1757$-$1854 were also carried out with the GBT, located in Green Bank, West Virginia, over the course of two observing semesters between 2024 and 2025 (see also Table~\ref{tab:observations}). Observations were carried out using the VEGAS backend in Vegas Pulsar Mode (VPM) and all observations made use of the L-band receiver at a centre frequency of 1500 MHz with an 800 MHz bandwidth. Each coherently dedispersed observation was approximately 5.15 hours long and recorded in full-Stokes polarisation mode. Once raw data was obtained, the GBT dataset was folded using \textsc{dspsr} \citep{vanStraten_Bailes2010} to produce standard archive files with 256 bins, 512 frequency channels and 6 minute sub-integration lengths. The number of frequency channels were further scrunched to 4 for timing analysis using the \textsc{psrchive} \citep{HotanPSRCHIVE2004} utility \textit{pam}.
A Chebyshev polynomial predictor, obtained with \textsc{tempo2} \citep{HobbsTempo22006}, was also used in the folding process to predict the pulse-phase of PSR~J1757$-$1854 using both frequency- and time-domain coefficients in order to mirror the folding routine used on the GBT data in \cite{cameron2023new}. A typical S/N of 108 was achieved for these data per orbit. 

The timing analysis (described in Sec.~\ref{Sec:TimingAnalysis}) was performed partly in the ilifu high-performance computing facility, hosted at the University of Cape Town ICTS and operated by IDIA on behalf of a consortium of universities and research organisations in South Africa and partly in the Hercules computing cluster of the Max Planck Insitute for Radio Astronomy, Germany.

\begin{table}
\centering
\caption{Timing parameters for PSR~J1757$-$1854. Numbers in parentheses are 1$\sigma$ uncertainties on the last digit(s).}
\label{tab:J1757timingparameters}
\begin{tabular}{ll}
\hline
\hline
Parameter & Value \\
\hline
\multicolumn{2}{c}{\textbf{Astrometric \& Spin parameters}} \\
\hline
Right Ascension, $\alpha$ (J2000) & 17:57:03.7821836(11) \\
Declination, $\delta$ (J2000) & $-18$:54:03.37288(25) \\
Proper motion in RA, $\mu_\alpha$ (mas\,yr$^{-1}$) & $-4.46(4)$ \\
Proper motion in DEC, $\mu_\delta$ (mas\,yr$^{-1}$) & $-2.8(6)$ \\
Spin frequency, $\nu$ (Hz) & 46.5176158883935(21) \\
Spin frequency derivative, $\dot{\nu}$ (s$^{-2}$) & $-5.68353(32)\times10^{-15}$ \\
Spin frequency 2nd derivative, $\ddot{\nu}$ (s$^{-3}$) & $-4.1(16)\times10^{-27}$ \\
Spin frequency 3rd derivative, $\dddot{\nu}$ (s$^{-4}$) & $-1.07(6)\times10^{-33}$ \\
Spin frequency 4th derivative, $\ddddot{\nu}$ (s$^{-5}$) & $-1.68(9)\times10^{-41}$ \\
Reference epoch (MJD) & 60000.104420185 \\
\hline
\multicolumn{2}{c}{\textbf{Dispersion parameters}} \\
\hline
Dispersion measure, DM (pc\,cm$^{-3}$) & 378.194(1) \\
DM derivative, DM1 (pc\,cm$^{-3}$\,yr$^{-1}$) & $-0.0168(4)$ \\
DM 2nd derivative, DM2 (pc\,cm$^{-3}$\,yr$^{-2}$) & $-0.0053(1)$ \\
\hline
\multicolumn{2}{c}{\textbf{Binary parameters}} \\
\hline
\multicolumn{2}{l}{\emph{Keplerian parameters}} \\
Orbital period, $P_\mathrm{b}$ (d) & 0.183537823671(2) \\
Epoch of periastron, $T_0$ (MJD) & 60004.142219871(15) \\
Projected semi-major axis, $x$ (s) & 2.2378066(9) \\
Longitude of periastron, $\omega$ (deg) & 344.70107(4) \\
Eccentricity, $e$ & 0.605817(2) \\
\hline
\multicolumn{2}{l}{\emph{Post-Keplerian parameters}} \\
Orbital period derivative, $\dot{P}_\mathrm{b}$ & $-5.282(2)\times10^{-12}$ \\
Periastron advance, $\dot{\omega}$ (deg\,yr$^{-1}$) & 10.364986(8) \\
Einstein delay parameter, $\gamma$ (s) & $0.0035931(33)$ \\
Orthometric amplitude, $h_3$ (s) & $4.97(12)\times10^{-6}$ \\
Orthometric ratio, $\varsigma$ & $0.911(6)$ \\
Relativistic deformation, $\delta_\theta$ & $5.0(23)\times10^{-6}$ \\
\hline
\multicolumn{2}{c}{\textbf{Mass Measurements (based on DDGR)}} \\
\hline
Total system mass, $M_\mathrm{tot}$ (M$_\odot$) &  2.7328884(36) \\
Pulsar mass, $m_{\rm p}$ (M$_\odot$) &   1.3384(2)\\
Companion mass, $m_{\rm c}$ (M$_\odot$) &  1.3945(2)\\
\hline
\multicolumn{2}{c}{\textbf{Derived parameters}} \\
\hline
Galactic longitude, $\ell$ ($\deg$) & 9.96610 \\
Galactic latitude, $b$ ($\deg$) & 2.87698 \\
Ecliptic longitude, ($\deg$) & 269.30318 \\
Ecliptic latitude, ($\deg$) & 4.53658 \\
Spin period, $P_\mathrm{p}$ (s) & 0.0214972324118936(9)\\
Spin period derivative, $\dot{P_\mathrm{p}}$ (s/s) & $2.62653(1) \times 10^{-18}$ \\
\hline
 \\
\end{tabular}
\end{table}


\section{Timing Analysis}
\label{Sec:TimingAnalysis}
The timing solution for PSR~J1757$-$1854 based on 6 yrs of data obtained with the GBT and the Murriyang (previously Parkes) radio telescope was presented by \cite{cameron2023new}. We extend the total timing baseline to 9 yrs, by including additional observations from MeerKAT along with new data from the GBT. In this section, we describe the methodology adopted to derive the updated timing solution. 

The templates used to calculate the ToAs for the MeerKAT data in two different bands were generated in the following steps. A high S/N profile was obtained, for each of the two frequency bands of the MeerKAT data, L-band and S1-band, by adding all the data recorded at these respective bands using the \textit{psradd} function in \textsc{psrchive}. These high S/N profiles were then split into 8 frequency channels using the function \textit{psrsplit} in \textsc{psrchive} \citep{HotanPSRCHIVE2004}. Hereafter, the \textsc{psrchive} function \textit{paas} was used to generate a noise-free profile for the lowest frequency channel by fitting 3 von-Mises functions to the pulse profile. The final model was saved and then applied to the next frequency channel to obtain a noise-free profile, by allowing the width and the amplitudes of the components to vary but not their centroid. This procedure was iteratively repeated up to the highest frequency channel profile in both bands. Finally the template profiles at different frequency channels were added using \textit{psradd} to obtain a frequency-resolved noise-free template in the two frequency bands. This method was also used  to generate the template for the new data from the GBT with 4 frequency channels (reduced from the raw 512 channels as mentioned in Sec.~\ref{Sec:Obs-datareduction}).  
These frequency-resolved templates were used to calculate the ToAs for the three different datasets using \textit{pat} within \textsc{psrchive}. We used the Fourier-domain Markov chain Monte Carlo algorithm (FDM) to generate the ToAs.

The timing analysis was carried out using the pulsar timing software package \textsc{tempo2} \citep{HobbsTempo22006}. 
The MeerKAT and GBT ToA datasets were combined with the earlier observations presented in \cite{cameron2023new} by introducing phase offsets (\rm{JUMPs}) between different subsets of the data. This step is necessary because the ToAs for these datasets were generated using different template profiles, which can introduce arbitrary phase offsets. Initial timing solutions were obtained using \textsc{tempo2} by fitting the DDH binary model \citep{freire2010orthometric}. These preliminary timing solutions were used to sample the timing model and determine noise models in the combined dataset. We considered three different noise models of increasing complexity. The first model  (WN) consists of basic white noise only, which accounts for radiometer noise, instrumental systematics, and intrinsic pulse phase jitter. In this framework, the ToA uncertainties are rescaled using two parameters: \rm{EFAC}, a multiplicative factor applied to the nominal ToA uncertainties, and \rm{EQUAD}, an additional variance term added in quadrature. The effective variance of a ToA is therefore given by
\begin{equation}
\label{wn}
\sigma^2 = \mathrm{EFAC}^2 \, \sigma_r^2 + \mathrm{EQUAD}^2\text{,}
\end{equation}
where $\sigma_r$ is the uncertainty associated with each ToA. It is important to note here that there could be other contributions to white noise which have not been incorporated in this model. The second model (WN+RN) includes the white noise model and a red noise model to account for the intrinsic pulsar spin noise. This red noise is modeled as a stationary Gaussian process with the power spectral density described by a power-law of the form
\begin{equation}
\label{tn}
\mathcal{P}(f) = \frac{A_{\mathrm{red}}^2}{12\pi^2} \left( \frac{f}{f_{\mathrm{yr}}} \right)^{-\gamma}\text{,}
\end{equation}
with $A_{\mathrm{red}}$ as the dimensionless red-noise amplitude (defined at a reference frequency $f_{\mathrm{yr}} = 1\,\mathrm{yr}^{-1}$) and $\gamma$ the spectral index. The third model (WN+RN+DM) further incorporates DM variations, which arise from changes in the integrated electron column density along the line of sight. These DM variations are modeled as an additional red-noise process with a similar power-law spectrum, but with a chromatic dependence on the observing frequency ($\propto \nu^{-2}$), thereby affecting ToAs differently across frequency bands.

The Bayesian parameter estimation and model selection were performed using the \textsc{temponest} framework \citep{LentatiTemponest2014}, which implements a nested sampling algorithm to simultaneously infer timing parameters and noise model parameters while computing the Bayesian evidence for each model. This enables a statistically robust comparison between competing noise models. Finally, since the ToAs were generated using different template profiles including different reference DMs, we included \rm{FDJUMPDM} parameters \citep{Bhat_etal2026} in the timing model to account for systematic DM offsets between datasets.

\begin{figure}
  \centering
  \includegraphics[scale=.315, trim={1.35 0.5 1.0 0.5},clip]{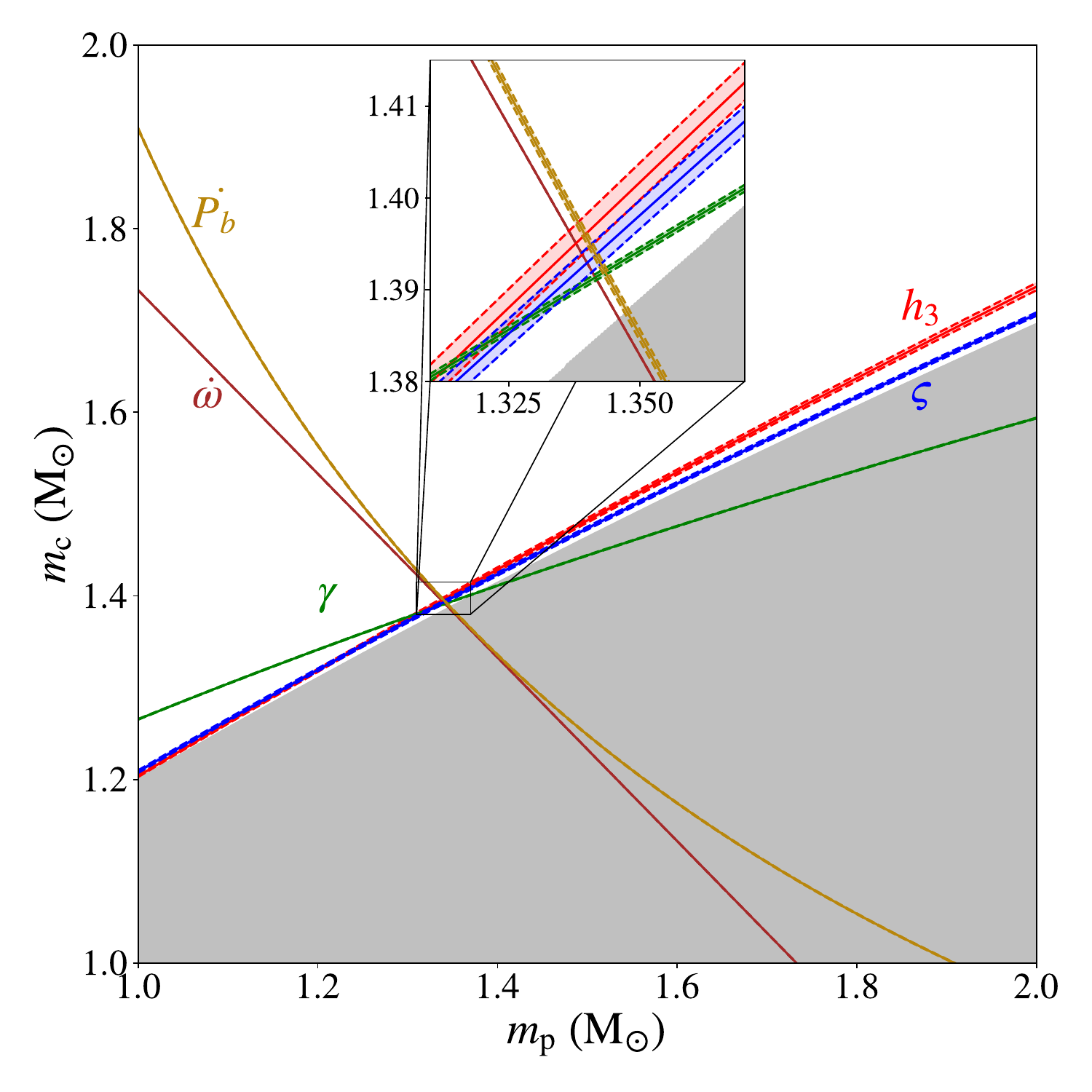}
  \caption{Mass-mass diagrams for PSR~J1757$-$1854 showing constraints from the measured post-Keplerian parameters. Each curve represents the locus of companion mass $m_\mathrm{c}$ versus pulsar mass $m_\mathrm{p}$ permitted by a given observable, assuming GR. The $1-\sigma$ uncertainties of the regions consistent with the measured 5 PK parameters are shown in shaded regions in brown ($\dot{\omega}$), golden ($\dot{P}_\mathrm{b}$), green ($\gamma$), red ($h_3$), and blue ($\varsigma$). As evident, the $\dot{P}_\mathrm{b}$ and $\gamma$ curves are divergent from the common intersection of the other parameters. The divergence of $\dot{P}_\mathrm{b}$ is due to the uncertainty in the known distance of the pulsar as discussed in \citet{cameron2023new}. 
  The divergence of gamma arises because we do not account for orbital deformation, $\delta_{\theta}$. In Figure~\ref{fig:mm_J1757DDHcorrectGamma}, we present a mass-mass diagram which accounts for this effect.}
  \label{fig:mm_J1757DDHwrongGamma}
\end{figure}


\section{Results and discussions}
\label{Sec:ResultsDiscussion}

We present here the results of the timing analysis. The noise analysis performed using \textsc{temponest} favoured the WN model with a Bayes factor of 3087 and 2682 as compared to the WN+RN and WN+RN+DM models respectively. All further analysis have been done by fixing the WN model parameters. We do not use WN+RN and WN+RN+DM models for any further analysis. 

\subsection{Timing Parameters}
The timing solution derived from our analysis of the ToA dataset is presented in Table~\ref{tab:J1757timingparameters}, and the post-fit timing residuals for the full combined dataset are shown in Fig.~\ref{fig:timing-residuals}. The timing residuals have a weighted rms of w$_{\rm rms}$ = 28.6$~\mu$s.

The combination of the previous dataset with the high-sensitivity MeerKAT data and new data from GBT, extending the timing baseline to 9 yrs, has led to substantial improvements in the precision of essentially all timing parameters (see Table~\ref{tab:J1757timingparameters}) compared to the six-year solution presented in \cite{cameron2023new}. The astrometric parameters are now determined with significantly higher precision. The precision in proper motion in right ascension $\mu_\alpha$ has been improved by a factor of $\sim 3$, whereas $\mu_\delta$ has been improved by a factor of 2 and is now detected with $\sim 4.5 \sigma$ significance. The extended dataset requires the inclusion of higher-order spin frequency derivatives beyond $\dot{\nu}$ in the timing model, as listed in Table~\ref{tab:J1757timingparameters}. The precision achieved in measuring the five PK parameters, all of which were also detected by \cite{cameron2023new}, has improved significantly in the present work. The precision in the detection of the the rate of periastron advance $\dot{\omega}$ has improved by an order of magnitude. The orbital period derivative $\dot{P}_\mathrm{b}$, and the orthometric Shapiro delay parameters, $h_3$ and $\varsigma$, have improved in precision by approximately a factor of two. If we blindly fit the Einstein delay parameter $\gamma$ as in \cite{cameron2023new}, we also obtain an improved value, $\gamma = 3.5860(9)\, \rm ms$, with the uncertainty being $\sim1.6$ times smaller than in that previous work. 

The mass-mass diagram for PSR~J1757$-$1854, constructed using the PK parameter measurements from the DDH timing solution, is shown in Fig.~\ref{fig:mm_J1757DDHwrongGamma}. Assuming GR, each PK parameter defines a curve in the $m_\mathrm{p}$-$m_\mathrm{c}$ plane. The intersection of any two curves yields the individual component masses; if the component masses derived from all pairs of curves are consistent with each other, then GR passes this self-consistency test. The improved precision of the PK parameters in this work has substantially narrowed the allowed region in the mass-mass plane, yielding tighter constraints on both the pulsar mass $m_\mathrm{p}$ and the companion mass $m_\mathrm{c}$.

\begin{figure}
  \centering
  \includegraphics[scale=.2445, trim={1.2 0 0 0},clip]{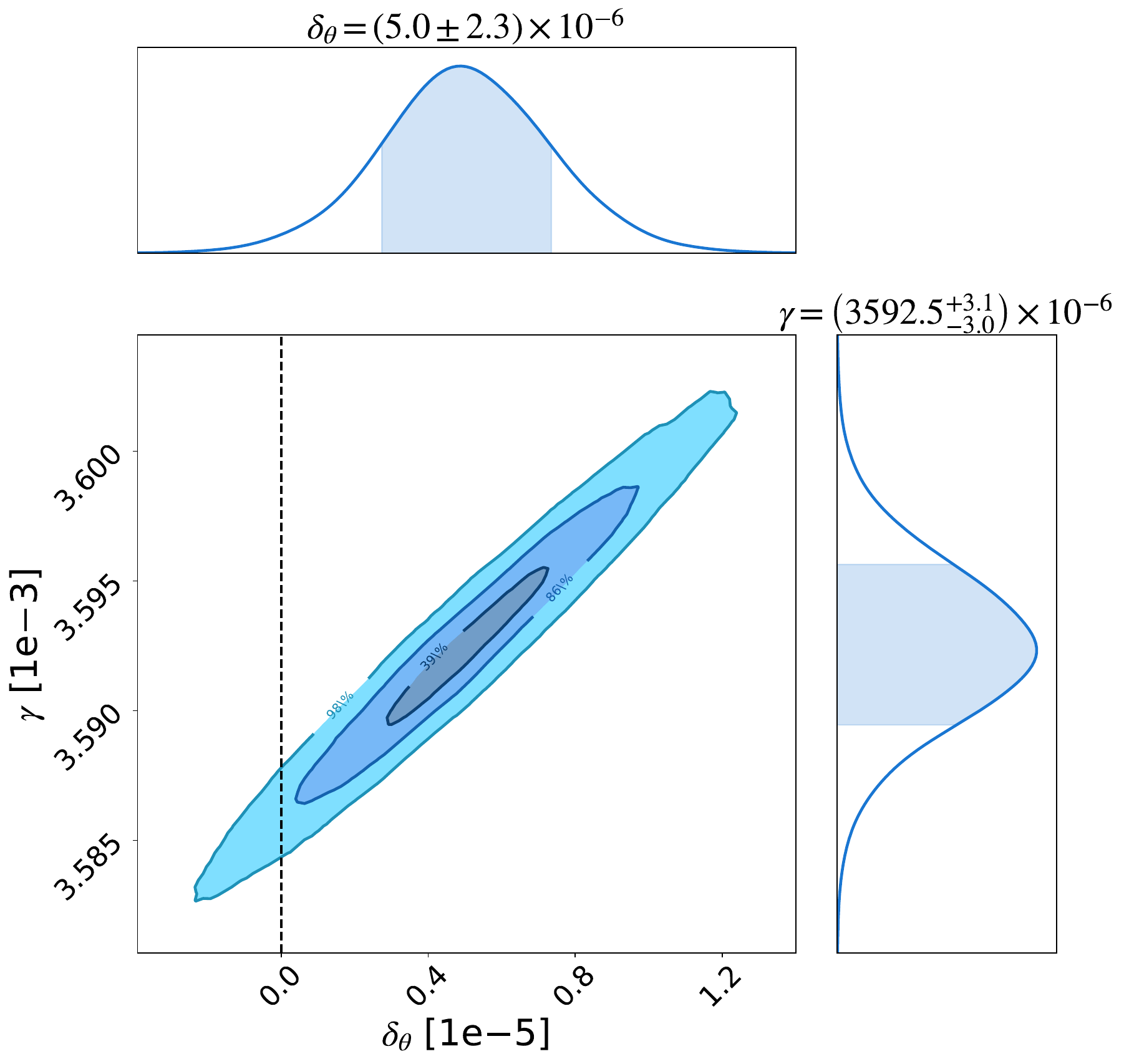}
  \caption{Joint posterior distribution of the post-Keplerian parameters $\delta_\theta$ and $\gamma$ for PSR~J1757$-$1854, sampled within the DDH timing framework of \textsc{tempo2}. The 2D contours reveal a strong positive correlation between the two parameters. The dotted line in the 2D distribution represents the $\delta_\theta=0$ line illustrating the deviated corresponding $\gamma$ value for which the mass-mass constraints curve is shown in Fig.~\ref{fig:mm_J1757DDHwrongGamma}.} 
  \label{fig:J1757Dtheta-Gamma}
\end{figure}

However, two notable features emerge from inspection of the mass-mass diagram. First, the GR curve of the orbital period derivative $\dot{P}_\mathrm{b}$ (in gold) does not pass through the intersecting region defined by the other PK parameters, showing a significant offset. This is similar to the offset observed by \cite{cameron2023new}. The offset is due to the uncertainty in the distance to this pulsar, which results in an uncertainty in the estimate of the kinematic corrections to $\dot{P}_\mathrm{b}$. Second, the value of $\gamma$ derived in our initial timing solution shows a notable deviation from the intersecting region defined by the remaining PK parameters ($\dot{\omega}$, $h_3$, and $\varsigma$). This systematic offset, which was already seen by \cite{cameron2023new} (but not with such high significance) is not attributable to measurement error and instead strongly suggests the presence of an unmodeled contribution to the timing residuals that gets absorbed onto the $\gamma$ parameter in the DDH fit. As we discuss next, this deviation is naturally explained by the relativistic angular deformation parameter $\delta_\theta$, which is correlated with $\gamma$ and was not included in the initial DDH timing solution, nor in the fit by \citet{cameron2023new}. The fact that we now take this correlation into account is the reason for the apparent increase in the uncertainty of $\gamma$ presented in Table \ref{tab:J1757timingparameters} compared to \citet{cameron2023new}.

\begin{figure}
  \centering
  \includegraphics[scale=.315, trim={1.35 0.5 1.0 0.5},clip]{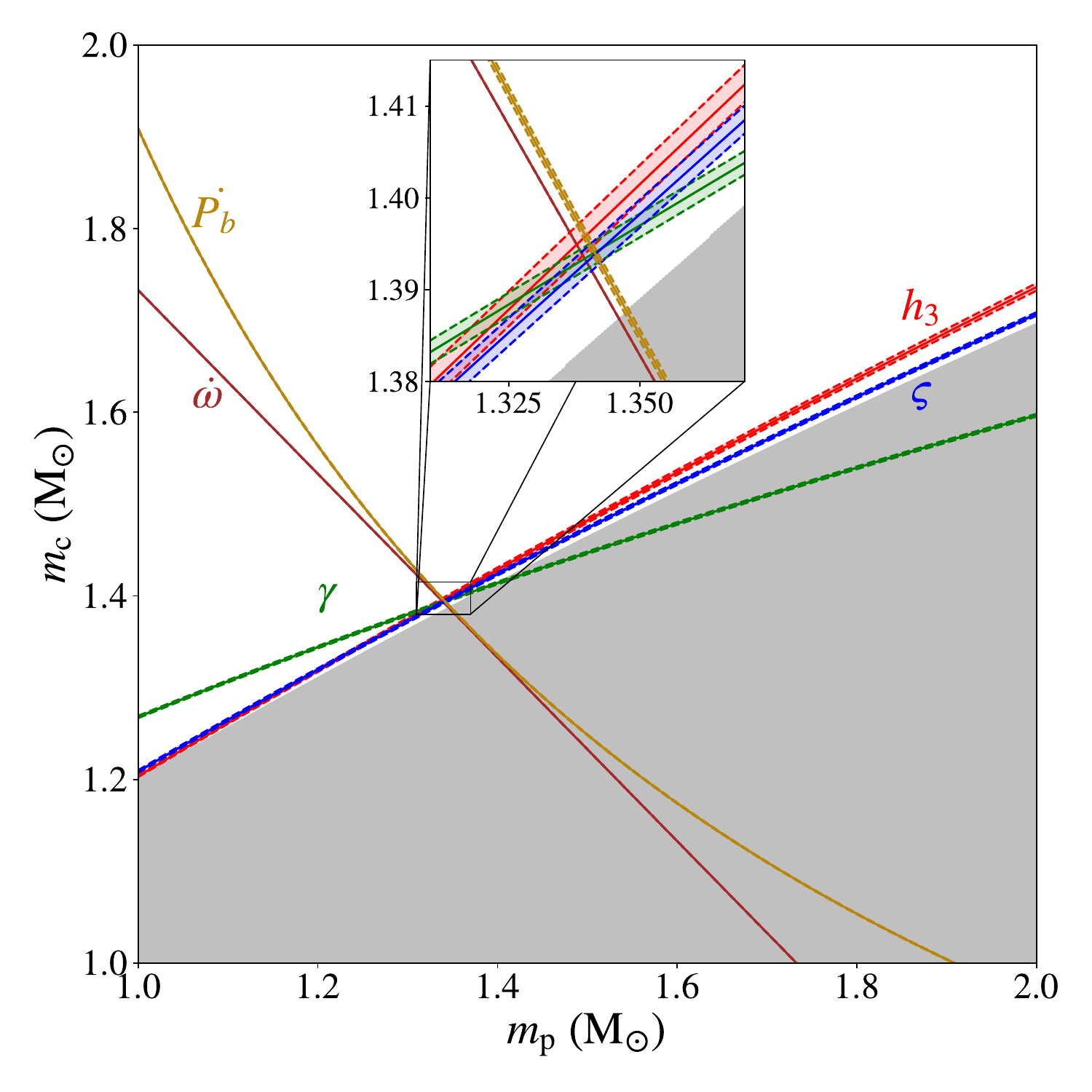}
  \caption{Mass-mass diagram for PSR~J1757$-$1854 constructed using the DDH timing solution and the value of $\gamma$ obtained from sampling it together with  $\delta_{\theta}$. The divergence seen in $\gamma$ in Fig.~\ref{fig:mm_J1757DDHwrongGamma} has been resolved here due to the presence of $\delta_{\theta}$.}
  \label{fig:mm_J1757DDHcorrectGamma}
\end{figure}

\subsection{Relativistic deformation of the orbit}\label{SubSec:dtheta}

In the DD model, the relativistic deformation of the orbit is described by two parameters: the radial deformation $\delta_r$ and the angular deformation $\delta_\theta$. These parameters describe purely periodic relativistic corrections to the Keplerian motion. They describe the fractional offset between the time eccentricity of the DD model ($e$ in this paper) and the radial and angular eccentricities respectively. For this reason, they contain a contribution arising purely from time dilation (see, e.g., eqs. (3.12) and (3.13) in \citealt{damour1992strong}).\footnote{In alternative theories of gravity, $\delta_r$ and $\delta_\theta$ may also receive contributions from variations in the pulsar’s moment of inertia arising from changes in the local gravitational constant \citep{damour1992strong}.}
The small difference $\delta_\theta - \delta_r$ leads to a deviation of the pulsar’s orbit from an (rotating) elliptical shape when the orbit is expressed in harmonic coordinates.  
In GR, both $\delta_r$ and $\delta_\theta$ are functions of the component masses and orbital period, making them in principle testable predictions of the theory. In practice, $\delta_\theta$ is the more accessible of the two parameters. As discussed in detail by \citet{damour1986general}, both $\delta_r$ and $\delta_\theta$ are degenerate with classical parameters. Over longer timescales, $\delta_\theta$ can in principle be measured but due to the correlation of $\delta_r$ with the pulsar phase it is essentially unmeasurable. 

In this work, we report the detection of the relativistic angular deformation parameter $\delta_\theta$ for PSR~J1757$-$1854. This makes PSR~J1757$-$1854 only the third DNS system in which $\delta_\theta$ has been detected, following the Hulse-Taylor binary, B1913+16 \citep{Weisberg_Huang2016} and the double pulsar PSR~J0737$-$3039A/B \citep{kramer2021}, where it was measured at $\sim 1.5\,\sigma$. For PSR~J1757$-$1854, we obtain a detection at $\sim 2\,\sigma$ significance in an observing time span of only 9 yrs. A critical aspect of measuring $\delta_\theta$ is its strong correlation with $\gamma$, as has been demonstrated by \cite{kramer2021}. It is therefore essential to sample $\delta_\theta$ and $\gamma$ simultaneously in order to obtain unbiased constraints on both.

We adapted \textsc{tempo2} to be sensitive to the $\delta_\theta$ parameter, enabling joint sampling of $\delta_\theta$ and $\gamma$ within the timing model. We employed the affine-invariant ensemble sampler \textsc{emcee} \citep{Foreman-Mackey_etal2019} for this purpose, while simultaneously fitting for all other parameters in the timing model. The resulting $\chi^2$ posterior distribution is shown in Fig.~\ref{fig:J1757Dtheta-Gamma}, and clearly illustrates the strong correlation between the two parameters. When $\gamma$ is held fixed at the value obtained from the joint sampling and $\delta_\theta$ is sampled independently, the detection significance increases substantially to $\sim 7\,\sigma$. This improvement arises because fixing $\gamma$ removes the dominant source of degeneracy, allowing $\delta_\theta$ to be constrained more tightly by the data. Using this refined value of $\gamma$ together with the remaining timing parameters from the DDH solution, we construct an updated mass-mass diagram, shown in Fig.~\ref{fig:mm_J1757DDHcorrectGamma}. The new constraints on the pulsar mass $m_\mathrm{p}$ and companion mass $m_\mathrm{c}$ derived from $\gamma$ agree better with the constraints derived from other PK parameters than those derived from the value of $\gamma$ obtained without $\delta_\theta$ (Fig.~\ref{fig:mm_J1757DDHwrongGamma}), providing further confidence in the robustness of the detection and the self-consistency of the timing solution within GR. 

Beyond its role as a test of GR, the measured value of $\delta_\theta$ provides an independent means of constraining the spin-orbit geometry of PSR~J1757$-$1854. The observed value of $\delta_\theta$ receives contributions both from the intrinsic relativistic orbital deformation and from a correction term, $\epsilon_A$ due to the aberration effect that depends on the orientation of the pulsar spin axis relative to the orbital plane \citep{damour1992strong}. Specifically, the observed $\delta_\theta$ can be decomposed as:
\begin{equation}
\delta_{\theta}^\text{obs} =
\delta_{\theta}^\text{intrinsic} - \epsilon_A\text{,}
\end{equation}
where the correction term is defined as
\begin{equation}
\epsilon_A =
\frac{A}{x}
\end{equation}
with
\begin{equation}
A = -
\left( \frac{T_\odot^{1/3}}{(2\pi)^{2/3}} \right) \left(\frac{P_\mathrm{p}}{P_\mathrm{b}^{1/3} (1-e^2)^{1/2}} \right) \left( \frac{m_\mathrm{c}}{(m_\mathrm{p} + m_\mathrm{c})^{2/3}}\right) 
\left( \frac{\sin \eta}{\sin \lambda} \right)\text{.} 
\end{equation}
Here, $P_\mathrm{p}$ is the pulsar spin period; $T_{\odot} \equiv ({\cal G M})_{\odot}^{\rm N} / c^3
= 4.925490947641266978...\,\upmu \rm s$ is an exact quantity, the nominal solar mass parameter
$({\cal G M})_{\odot}^{\rm N}$ in time units \citep{Prvsa_etal2016}\footnote{In the equations containing $T_{\odot}$, the mass values are adimensional, expressing the ratio $Gm / ({\cal G M})_{\odot}^{\rm N}$, where $m$ is the corresponding mass in mass units. Explicit mass values in the text are followed by the symbol M$_{\odot}$ to indicate that they are multiples of the solar mass parameter.}; 
$\eta$ and $\lambda$ are the geodetically precessing polar angles of the pulsar spin axis with respect to the line of nodes and the line of sight, respectively (see \citealt{damour1992strong} for their definition and geometrical meaning). 
The correction term $\epsilon_A$ therefore encodes information about the three-dimensional orientation of the spin axis and evolves in time as the spin axis precesses geodetically around the orbital angular momentum vector. As noted by \citet{Weisberg_Huang2016}, a measurement of $\delta_\theta$ can help constrain the spin orientation of the pulsar through its contribution to $\epsilon_A$. 

\begin{figure}
  \centering
  \includegraphics[scale=.365, trim={1.25 0.5 1.0 0.5},clip]{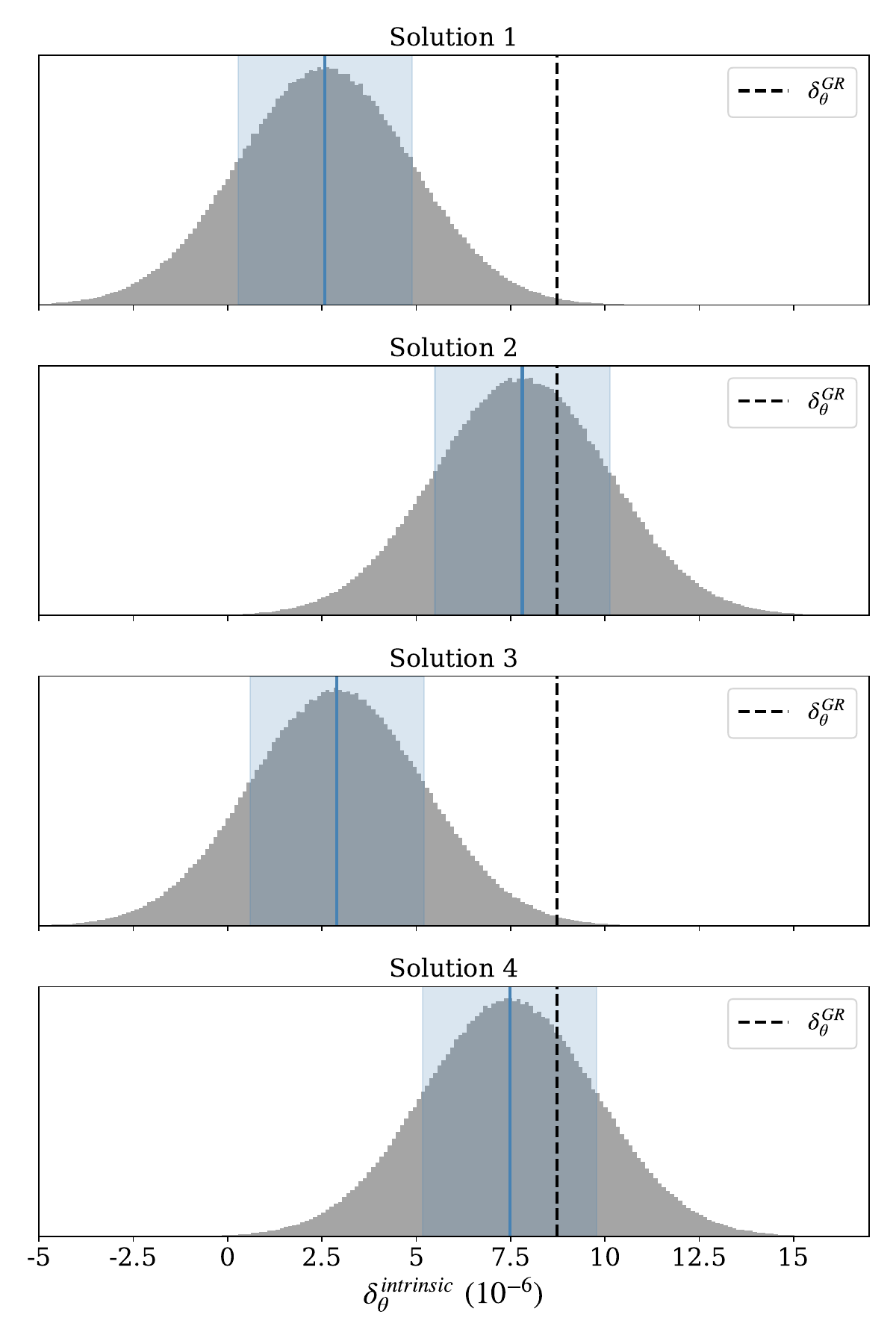}
  \caption{Posterior distributions of the intrinsic relativistic orbital deformation parameter $\delta_\theta^\mathrm{intrinsic}$ for each of the four geometric spin-orbit solutions of PSR~J1757$-$1854, as shown in Table~3 of \protect\cite{cameron2023new}. The blue vertical line and shaded region indicate the median and $1-\sigma$ credible interval for each solution, while the dashed black line marks the GR prediction $\delta_\theta^\mathrm{GR}$
 computed from the measured component masses and orbital parameters. Solutions 1 and 3 are inconsistent with the GR prediction, as their posterior distributions are significantly offset from $\delta_\theta^\mathrm{GR}$ value whereas Solutions 2 and 4 remain consistent, narrowing the viable spin-orbit geometries for this system.}
  \label{fig:dtheta-distributions}
\end{figure}
\begin{figure*}
  \centering
  \includegraphics[scale=.315, trim={1.0cm 0.6cm 0.7cm 0.5cm},clip]{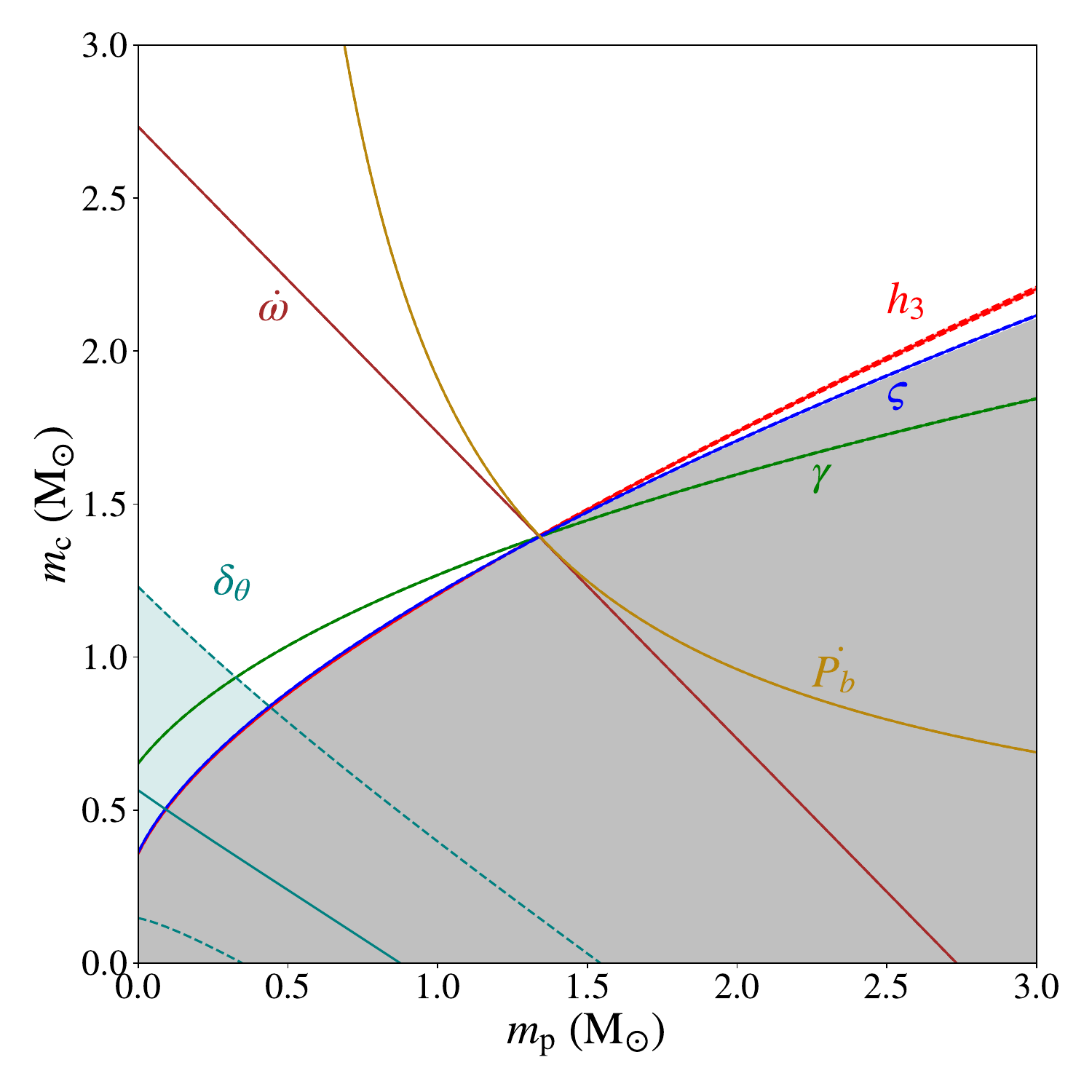}
  \hspace{0.5cm}
  \includegraphics[scale=.315, trim={1.0cm 0.6cm 0.7cm 0.5cm},clip]{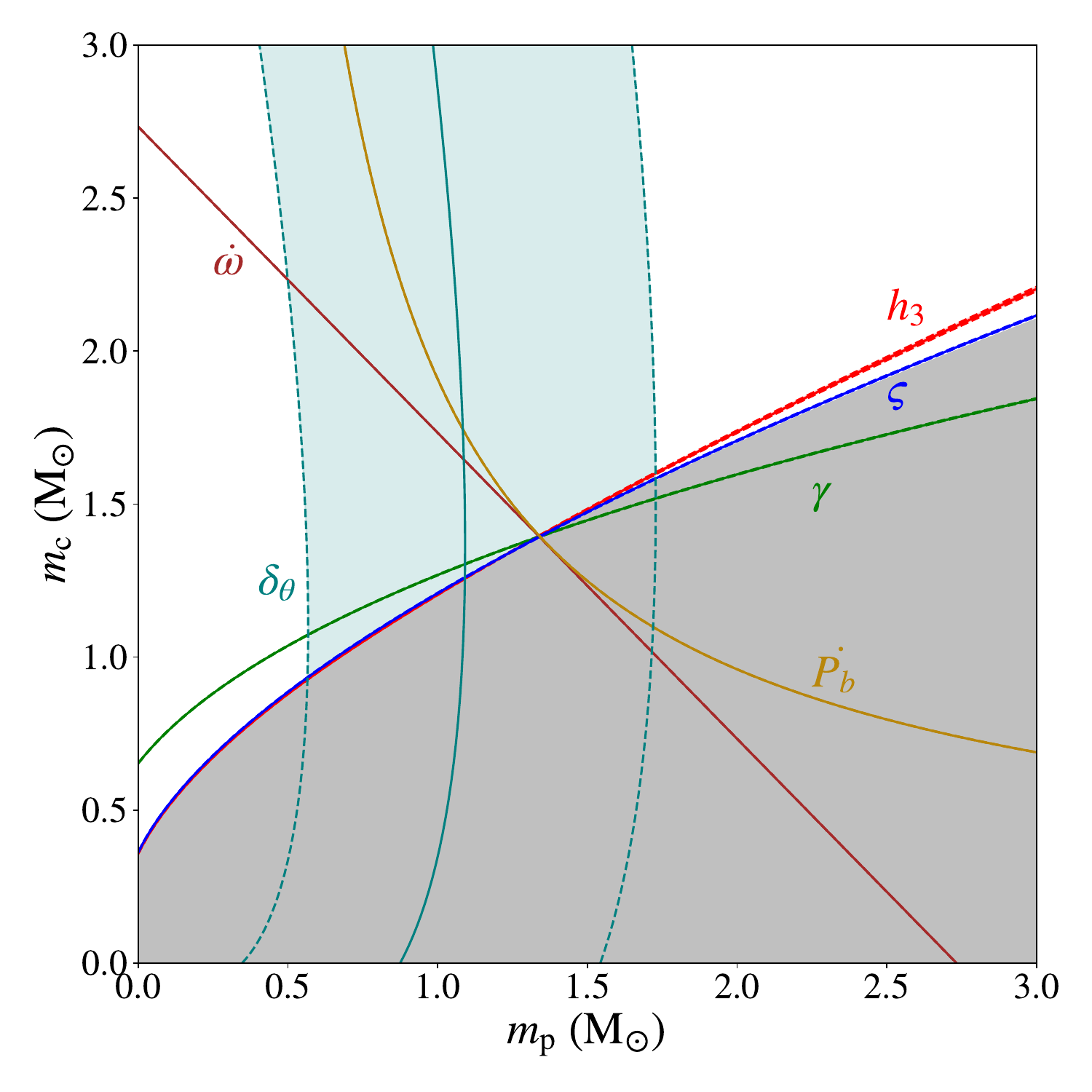}
  \caption{Mass-mass diagrams for PSR~J1757$-$1854 showing constraints, including that obtained with the relativistic deformation parameter $\delta_\theta$ (shown in teal with the shaded region indicating the $1-\sigma$ bounds). \textit{Left:} Constraints incorporating the spin-orbit geometry correction $\epsilon_A$ evaluated under Solution~1. \textit{Right:} Same, but under Solution~2. The markedly different width and position of the $\delta_\theta$ constraint band between the two panels reflects the distinct spin-axis orientations of each geometric solution, with Solution~2 yielding a $\delta_\theta$ band that passes through the region of mutual consistency, while Solution~1 does not. It is important to note here that a fixed value of the angles $\eta$ and $\lambda$ have been used for the full mass-mass plane in order to plot the $\delta_\theta$ constraints curve.}
  \label{fig:mmJ1757-withDtheta}
\end{figure*}

In GR, the purely relativistic (intrinsic) contribution to $\delta_\theta$ is determined entirely by the component masses and the orbital period:
\begin{equation}
     \delta_{\theta}^\text{GR} = \left( \frac{2 m_\mathrm{c}^2 + 6 m_\mathrm{p} m_\mathrm{c} + \frac{7}{2} m_\mathrm{p}^2}{(m_\mathrm{p} + m_\mathrm{c})^{4/3}} \right) 
\left( T_\odot n_\mathrm{b} \right)^{2/3},
\end{equation}
where $n_\mathrm{b} \equiv 2 \pi / P_\mathrm{b}$ is the orbital angular frequency of the system.
Given the well measured component masses of PSR~J1757$-$1854, this yields a precise theoretical prediction against which the observed $\delta_\theta$ can be compared.

\cite{cameron2023new} identified four distinct geometric solutions for the spin-orbit orientation of PSR~J1757$-$1854 (see Table 3 of \citealt{cameron2023new}). These are characterised by different values of the misalignment angle $\delta$ (the angle between the pulsar spin axis and the orbital angular momentum vector) and the reference precession phase $\Phi_{\mathrm{SO},0}$. The polar angles $\eta(t)$ and $\lambda(t)$ required to evaluate the aberration term, $\epsilon_A$ for each solution can be computed from these geometric parameters using the relations:
\begin{eqnarray}
  \cos\lambda(t) & = & \cos\delta\cos i - \sin\delta \sin i \cos \Phi_\text{SO}(t)
  \label{eqn:lambda}\\
  \cos\eta(t) \sin \lambda(t) & = & \sin\delta \sin\Phi_\text{SO}(t) \\
  \cos\delta & =& \cos\lambda(t) \cos i - \sin i \sin\lambda(t) \sin\eta(t)\text{,}
\end{eqnarray}
where the precession phase evolves as
\begin{equation}
\Phi_\text{SO}(t) = \Omega_\text{GP} (t-t_0) +\Phi_{\text{SO},0}\text{,}
\end{equation}
relative to the reference value $\Phi_{\text{SO},0} = \Phi_\text{SO}(t_0)$. The rate of geodetic precession $\Omega_\text{GP}$ is given by:
 \begin{equation}
     \Omega_{\mathrm{GP}} = \left(\frac{n_\mathrm{b}^{5/3}T_\odot^{2/3}}{2(1 - e^2)}\right)
\left(\frac{m_\mathrm{c} \left(4m_\mathrm{p} + 3m_\mathrm{c}\right)}{(m_\mathrm{p} + m_\mathrm{c})^{4/3}}\right)\text{,}
 \end{equation}
this allows us to account for the time evolution\footnote{This does not imply a timing model with the temporal evolution of $\eta$ and $\lambda$, instead their values have been translated from the values in \cite{cameron2023new} to the new epoch with the GR precession rate and subsequently used for further calculations.} of $\eta$ and $\lambda$.
With these values, we are able to calculate $A$, $\epsilon_A$ and then add the resulting value to $\delta_{\theta}^\mathrm{obs}$, obtaining the value of $\delta_{\theta}^\mathrm{int}$.

We performed a Monte-Carlo (MC) analysis to obtain the posterior distribution of $\delta_{\theta}^\mathrm{int}$ for each of the four geometric solutions found by \cite{cameron2023new}. The resulting distributions are shown in Fig.~\ref{fig:dtheta-distributions}. Comparing these with the GR prediction $\delta_\theta^\mathrm{GR}$, we find that the $\delta_{\theta}^\mathrm{intrinsic}$ values inferred under Solutions~1 and~3 are inconsistent with $\delta_\theta^\mathrm{GR}$, effectively ruling out these two orientations. Solutions~2 and~4 remain consistent with the GR expectation, reducing the set of viable geometric configurations for this system. This conclusion is further corroborated by the mass-mass diagrams shown in Fig.~\ref{fig:mmJ1757-withDtheta}, where the $\delta_\theta^\mathrm{int}$ constraint band passes through the region of mutual consistency with all other post-Keplerian parameters only for the viable solutions, while the band is significantly offset for the ruled-out orientations. This demonstrates that $\delta_\theta$ measurements provide a powerful and novel method for breaking degeneracies in the spin-orbit geometry of DNS systems, complementing constraints from pulse profile evolution studies. It is however noteworthy that while plotting the $\delta_\theta$ constraints curve in the mass-mass diagram, for practicality, a fixed value of the angles $\eta$ and $\lambda$ have been assumed, which in general is not true\footnote{It is important to emphasize that this is done only to plot the  $\delta_\theta$ constraints curve in the mass-mass diagram and does not impact the GR consistency test in Fig.~\ref{fig:dtheta-distributions}.}. This comes from the fact that we have used a fixed value of inclination angle and geodetic precession rate for the entire mass-mass plane. For a fully consistent analysis, we must use these values corresponding to all the points in the mass-mass plane and further use them for the full polarisation evolution studies, which is the beyond the scope of this paper.

Solutions~2 and~4 are also favored by the evolutionary model of PSR~J1757$-$1854: the excluded solutions require a supernova kick that has to be twice as large as the pre-supernova orbital velocity and aimed in a direction opposite to this orbital velocity. This means that not only must the supernova kick velocity be very large, but also relatively fine-tuned in direction and magnitude in order to produce a system like PSR~J1757$-$1854.

\subsection{Contributions to \texorpdfstring{$\dot{P}_\mathrm{b}$}{Pbdot}}

The observed value of orbital period derivative, $\dot{P}_\mathrm{b}$ receives contributions from several kinematic and astrophysical processes that must be carefully separated before a meaningful comparison with the GR prediction can be made. The excess $\dot{P}_\mathrm{b}$ can be decomposed as: 
\begin{equation}\label{eqn: pbdot contributions}
    \dot{P}_\text{b,exs} = \dot{P}_\text{b,obs} - \left(\dot{P}_\text{b,GR} + \dot{P}_\text{b,Gal} + \dot{P}_\text{b,Shk}\right)\text{,}
\end{equation}
where $\dot{P}_\mathrm{b,GR}$ is the intrinsic orbital decay due to gravitational wave emission as predicted by GR, $\dot{P}_\mathrm{b,Gal}$ is the contribution from the differential Galactic acceleration along the line of sight \citep{damour_taylor1991}, $\dot{P}_\mathrm{b,Shk}$  is the apparent period derivative arising from the proper motion of the system (the Shklovskii effect,  \citealt{shklovskii70}), 
and $\dot{P}_\mathrm{b,exs}$ is the residual excess term, which should be consistent with zero if GR is correct and all extrinsic contributions are properly accounted for. The Galactic acceleration term $\dot{P}_\mathrm{b,Gal}$ depends on the three-dimensional position of the pulsar within the Galactic potential and is computed using a model for the Milky Way mass distribution. This contribution is sensitive to the assumed Galactic potential model and introduces a systematic uncertainty that is difficult to reduce without an independent distance measurement. The Shklovskii term \citep{Edwards_etal2006} can be given as:
\begin{equation}\label{eqn: shklovskii}
    \dot{P}_\text{b,Shk} \simeq 2.10\times10^{-16} \left(\frac{d}{\text{kpc}}\right)\left(\frac{\mu_\text{T}}{\text{mas\,yr}^{-1}}\right)^{2}\left(\frac{P_\text{b}}{\text{d}}\right)\text{.}
\end{equation}
where $\mu_\mathrm{T}$ is the total proper motion of the system and $d$ is the distance to the pulsar. The proper motion of PSR~J1757$-$1854 has been measured to good precision in this work (see Table~\ref{tab:J1757timingparameters}), despite the fact that the pulsar has a low ecliptic latitude, which makes any measurements of positions and proper motions in the direction perpendicular to the ecliptic much less precise.

\begin{figure}
  \centering
  \includegraphics[scale=.39, trim={1.25 0 0.5 0},clip]{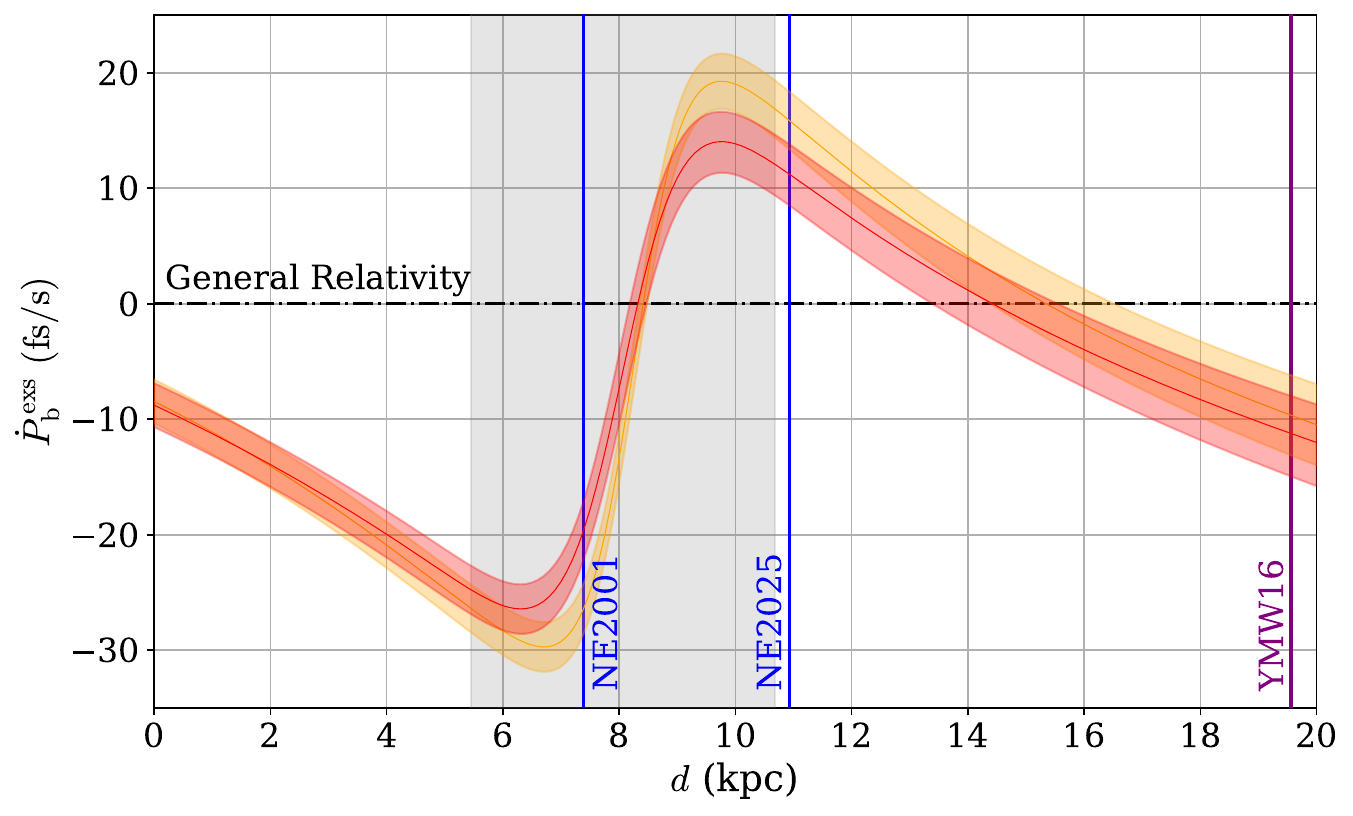}
  \caption{Excess orbital period derivative $\dot{P}_\mathrm{b,exs}$ for PSR~J1757$-$1854 as a function of distance $d$ (see Eq.~\ref{eqn: pbdot contributions}). The red and orange bands show the median and $1\sigma$ uncertainty of the estimation, where we have used two different models for the Galactic gravitational potential to calculate $\dot{P}_\mathrm{b}^\mathrm{Gal}$. The red band corresponds to the model of \protect\cite{McMillan_2017} and the orange to model I of \protect\cite{Irrgang_2013}. The dash-dotted horizontal line marks $\dot{P}_\mathrm{b,exs} = 0$, the value expected if GR fully accounts for the intrinsic orbital decay and all extrinsic contributions are correctly removed. The grey shaded region marks distances smaller than 3~kpc from the Galactic Centre, where the axisymmetric models of the Galactic potential used here are expected to be less reliable. The vertical blue and purple lines mark the distance estimates from the NE2001 \protect\citep{NE2001}, NE2025 \protect\citep{NE2025} and YMW16 \protect\citep{YMW2016} electron density models. It should be kept in mind that distances derived from electron density models may be subject to significant systematic uncertainties.}
  \label{fig:mm_J1757}
\end{figure}

The dominant source of uncertainty in the Shklovskii correction is therefore the poorly constrained distance $d$ to the system.
The Galactic electron density models provide a wide range of plausible distances to the pulsar (vertical lines in Figure~\ref{fig:mm_J1757}) \textbf{but the }uncertainties on these distances are not well characterised by these models. Within this broad range of possible distances, there are two distances ($\sim$8.4 and 13.5 to 15.5 kpc) where the $\dot{P}_\mathrm{b,exs}$ (given by the red and orange bands in  Fig.~\ref{fig:mm_J1757}) 
is zero, i.e. at those distances the observed $\dot{P}_\mathrm{b}$ is 1-$\sigma$ consistent with the GR prediction plus the extrinsic effects we calculate for those distances. These distances are certainly not excluded by the rather uncertain and mutually inconsistent DM distance estimates, i.e., we cannot rule out an agreement between the estimated intrinsic $\dot{P}_\mathrm{b}$ and the orbital period change due to the loss of orbital energy by gravitational waves as predicted by GR \citep{Peters_1964}. 
If a better estimate of the distance becomes available in the future, the observed $\dot{P}_\mathrm{b}$ could be converted into a more quantitative test of GR. 

\subsection{System masses}
In order to obtain direct estimates of the component masses within the framework of GR, we employed the DDGR binary model \citep{DDGR} using \textsc{tempo2}. The DDGR model uses the total mass $M_{\rm tot}$ and the companion mass $m_\mathrm{c}$ as the fundamental free parameters and takes all relativistic effects self-consistently into account using GR. Importantly, $\gamma$, $\delta_\theta$ and their correlation are fully taken into account in this analysis. This provides a direct and internally consistent determination of the component masses. Given the fact that the $\dot{P}_\mathrm{b}$ has a significant but poorly known contribution from kinematic effects, we have, in this fit, added a parameter to quantify this, XPBDOT. This has the result that the masses are not biased by the kinematic effects on $\dot{P}_\mathrm{b}$. The standard implementation of the DDGR model in \textsc{tempo2} is subject to a known numerical issue, which can cause instabilities during fitting when the mass parameter space is large. We therefore, adapted the DDGR model implementation in \textsc{tempo2} to circumvent this issue. The masses obtained from fitting the DDGR model are presented in Table~\ref{tab:J1757timingparameters}.

However, as we'll see next, these mass values, especially the total mass value, must be corrected because of second-order effects on $\dot{\omega}$, which are not taken into account in the DDGR model.

\subsection{Effects of higher order contribution to rate of advance of periastron, \texorpdfstring{$\dot{\omega}$}{omegadot}}
The relativistic advance of periastron, $\dot{\omega}$ is among the most precisely measured post-Keplerian parameters in PSR~J1757$-$1854, and its high precision makes it sensitive to contributions beyond the leading-order post-Newtonian (PN) term \citep{Robertson_1938, damour1985general, damour1986general, Damour_Schafer1987, Damour_Schafer1988}. In the standard pulsar timing framework, $\dot{\omega}$ is typically evaluated up to the 1PN approximation, which relates the measured value directly to the total system mass. However, at the level of precision achieved in this work, higher-order relativistic corrections become non-negligible and it is necessary to study their effects on the estimation of the total mass. The full expression for $\dot{\omega}$, including contributions up to the 2PN approximation and the Lense-Thirring (LT) spin-orbit coupling, can be written as:
\begin{equation}
    \dot{\omega}=\dot{\omega}^{\rm 1PN} + 
    \dot{\omega}^{\rm 2PN} + \dot{\omega}^{\rm LT}\text{,}
    \label{eq:om}
\end{equation}
where $\dot{\omega}^{\rm 1PN}$ and $\dot{\omega}^{\rm 2PN}$ are the 1PN and 2PN contributions to the periastron advance, and $\dot{\omega}^{\rm LT}$ is the LT contribution arising from the coupling between the spin angular momentum of the pulsar and the orbital motion. 
The leading-order 1PN contribution to the periastron advance is given by:
\begin{align}
    \dot{\omega}^{\rm 1PN} = \frac{3\,n_{\rm b}^{5/3}\,(T_{\odot}M_\mathrm{tot})^{2/3}}{(1-e^2)}.
    \label{eq:omdot1pn}
\end{align}
This expression depends only on the total mass $M_\mathrm{tot}$ and the orbital parameters $P_\mathrm{b}$ and $e$, both of which are measured with high precision. The 2PN correction introduces a mass-ratio dependence into the periastron advance and is given by:
\begin{align}
    \dot{\omega}^{\rm 2PN} = \frac{3\,n_{\rm b}^{7/3}\,(T_{\odot}M_\mathrm{tot})^{4/3}}{(1-e^2)^{-1}}\,f_{\rm O}\text{,}
    \label{eq:omdot2pn}
\end{align}
where the dimensionless function $f_{\rm O}$ encodes the dependence on the individual mass fractions and is defined as:
\begin{equation}
    \begin{split}
        f_{\rm O} = \frac{1}{1-e^2}&\left(\frac{39}{4}X_{\rm p}^2 + \frac{27}{4}X_{\rm c}^2 + 15\,X_{\rm p}X_{\rm c} \right) \\
        &- \left(\frac{13}{4}X_{\rm p}^2 + \frac{1}{4}X_{\rm c}^2 + \frac{13}{3}X_{\rm p}X_{\rm c} \right)\text{,}
    \end{split}
    \label{eq:fo}
\end{equation}
with $X_{\rm p} \equiv m_\mathrm{p}/M_\mathrm{tot}$ and $X_{\rm c} \equiv m_\mathrm{c}/M_\mathrm{tot}$ being the dimensionless mass fractions of the pulsar and companion, respectively. The LT contribution to $\dot{\omega}$ from the pulsar spin \citep{Barker_Connel1975, Damour_Schafer1988, FreireWex2024} is given by:
\begin{equation}
\dot{\omega}^{\rm LT} = \frac{A_\mathrm{p}}{\sin^2 i} \left[ (1 - 3\sin^2 i)\cos\delta - \cos i\,\cos\lambda \right]\text{,}
\label{eq:omLT}
\end{equation}
and $A_\mathrm{p}$ is given by:
\begin{equation}
A_\mathrm{p} = \left(\frac{4 + 3\,m_\mathrm{c}/m_\mathrm{p}}{2\,c^2\,M_\mathrm{tot}\,(1-e^2)^{3/2}}\right)\,2\pi\nu\,n_\mathrm{b}^2I_\mathrm{p}\text{,}
\label{eq:Ap}
\end{equation}
where $\nu$ is the spin frequency of the pulsar and $I_\mathrm{p}$ is its moment of inertia. The LT contribution therefore depends on the spin and geometric orientation of the pulsar, as well as on $I_\mathrm{p}$, which is sensitive to the neutron star equation of state. Since $I_\mathrm{p}$ is not independently known, we evaluate the LT contribution using the moment of inertia estimate derived in Appendix~B (Eq.~\ref{eq:MOI}), $I_\mathrm{p} = (1.39 \pm 0.08) \times 10^{45}\,\mathrm{g\,cm^2}$, together with the spin-orbit orientation parameters from the two geometric solutions in Table~3 of \cite{cameron2023new} (Solutions~2 and~4) that were not excluded by our measurement of $\delta_\theta$. The corrections to $\dot{\omega}$ have been estimated for this pulsar and the values are tabulated in Table~\ref{tab:omdot-contributions}.

\begin{table}
    \centering
        \caption{Individual contributions to the total periastron advance $\dot{\omega}$ for PSR~J1757$-$1854, expressed in degrees per year. The first two rows give the leading-order (1PN) and second-order (2PN) post-Newtonian contributions, which depend only on the component masses and Keplerian parameters. The final two rows give the Lense-Thirring contribution arising from coupling of the pulsar spin to the orbital angular momentum, evaluated under the two viable spin-orbit geometric solutions (Solutions~2 and~4) from \protect\cite{cameron2023new}.}
    \begin{tabular}{l l }
    \hline
    \hline
         &  $\dot{\omega}$   \\
         & ($\rm deg / yr$) \\
    \hline 
    
    1PN
    &    \,\! 10.364915(9)   \\
    2PN
    &    \;\: 0.000331110(4)  \\
    LT - Solution 2
    &    $-$0.00024(2)  \\
    LT - Solution 4
    &    $-$0.00019(2)   \\
    \hline
    \end{tabular}
    \label{tab:omdot-contributions}
\end{table}

\begin{figure}
  \centering
  \includegraphics[scale=.4, trim={0.5 0 0.5 0},clip]{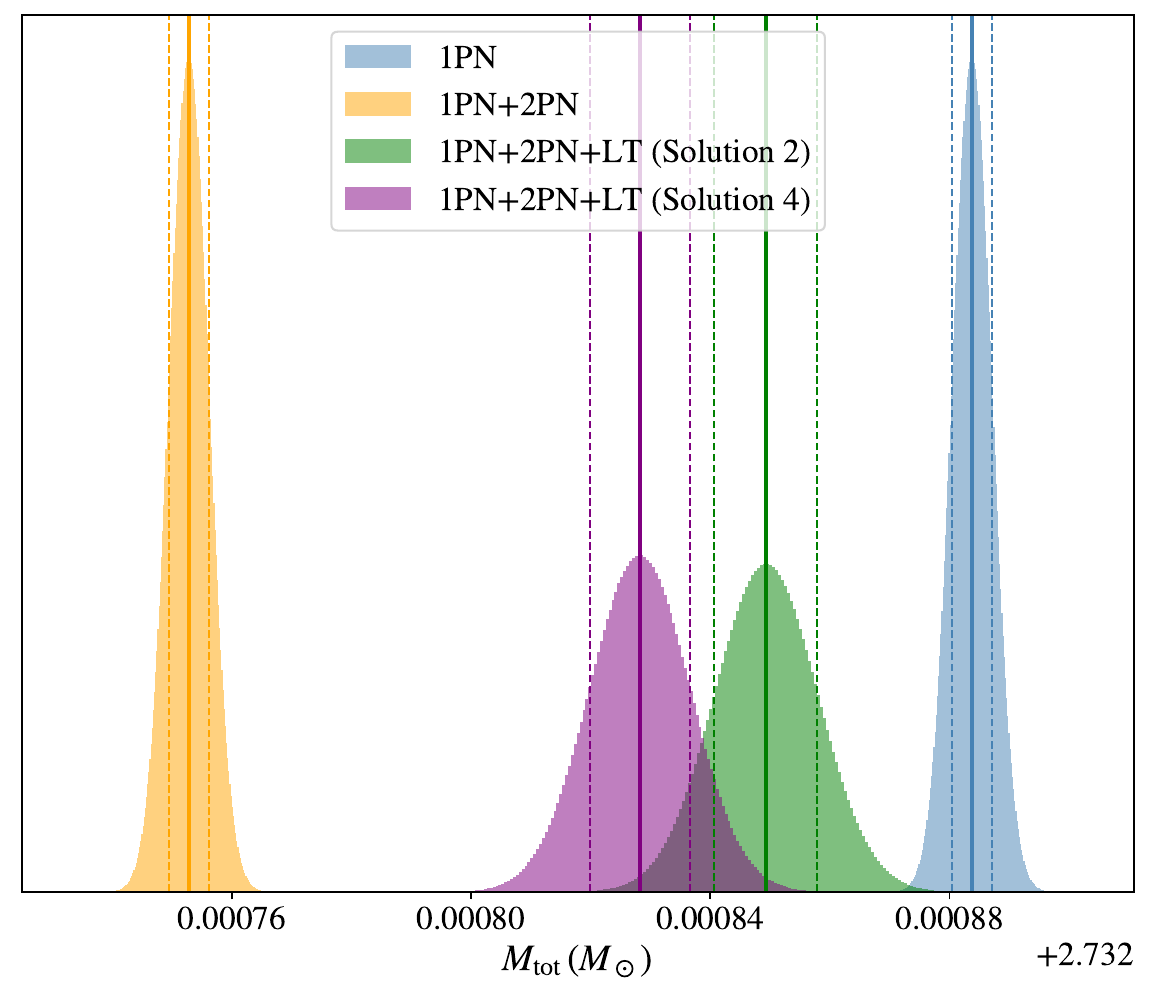}
  \caption{Distributions of the total mass $M_\mathrm{tot}$ for PSR~J1757$-$1854, derived from MC analysis of the periastron advance $\dot{\omega}$ with higher order contributing terms. The four cases shown are: the leading-order 1PN contribution alone (orange), the 1PN and 2PN contributions combined (yellow), and the full 1PN+2PN expression supplemented by the Lense-Thirring contribution evaluated for Solution 2 (green) and Solution 4 (purple) of the spin-orbit geometry. The dashed vertical lines mark the median of each distribution, and the shaded regions indicate the $1-\sigma$ credible intervals. The systematic shift in $M_\mathrm{tot}$ across the four cases demonstrates that higher-order relativistic and spin-orbit corrections to $\dot{\omega}$ have a measurable impact on the inferred total mass.} 
  \label{fig:Totalmass-omdot}
\end{figure}

We performed a MC analysis to estimate the total mass of the system due to all the contributions to $\dot{\omega}$ listed above. These values are presented in Table~\ref{tab:mass-omdot}. The total mass shows a significant difference, which suggests that the higher order contributions to $\dot{\omega}$ significantly impact the measurements of total mass. This is also seen in the Fig.~\ref{fig:Totalmass-omdot}, where the distribution of the total mass obtained from the MC analysis is presented. 

\begin{table}
    \centering
    \caption{Total system mass $M_\mathrm{tot}$ of PSR~J1757$-$1854 inferred from the periastron advance $\dot{\omega}$ after adding higher order terms. The variation in the inferred $M_\mathrm{tot}$ across the four cases, illustrates the importance of accounting for higher-order relativistic and spin-orbit corrections to $\dot{\omega}$ in this system.}
      
    \begin{tabular}{l l }
    \hline
    \hline
         &  $M_\mathrm{tot}$   \\
         & ($\rm M_{\odot}$) \\
    \hline     
    $\dot{\omega}^{\rm 1PN}$
    &   2.7328837(34) \\
    $\dot{\omega}^{\rm 1PN}+\dot{\omega}^{\rm 2PN}$
    &    2.7327528(34)   \\
    $\dot{\omega}^{\rm 1PN}+\dot{\omega}^{\rm 2PN}+\dot{\omega}^{\rm LT}$ (Solution 2)
    &   2.732849(9)     \\
    $\dot{\omega}^{\rm 1PN}+\dot{\omega}^{\rm 2PN}+\dot{\omega}^{\rm LT}$ (Solution 4)
    &    2.732828(8)     \\
    \hline
    \end{tabular}
    \label{tab:mass-omdot}
\end{table}


\section{Conclusion and Future Scope}
\label{Sec:ConclusionFuture}

\subsection{Summary of results}
In this work, we have presented the results of high-precision timing campaign for the highly relativistic double neutron star system PSR~J1757$-$1854, combining archival data from the Murriyang (previously Parkes) telescope and the Green Bank Telescope with new high-sensitivity observations from the MeerKAT radio telescope at both L-band and S1-band, as well as a much extended GBT data set.

The addition of MeerKAT data in particular has proved transformative, dramatically improving the quality of the combined dataset and enabling measurements that were previously unattainable. The nine-year timing solution presented here represents a significant advance over the six-year solution of \cite{cameron2023new}. The astrometric and spin parameters have been measured with substantially improved precision. Notably, the proper motion in right ascension, $\mu_\alpha$, has been measured with a factor of $\sim 3$ improvement in precision and is now detected at $\sim 4.5\sigma$ significance, providing a more reliable foundation for computing kinematic corrections to the observed orbital period derivative. The extended dataset has also necessitated the inclusion of higher-order spin frequency derivatives up to $\dddot{\nu}$ in the timing model, reflecting the increasing sensitivity of the solution to the long-term spin evolution of the pulsar. 

The post-Keplerian (PK) parameters previously measured by \cite{cameron2023new} are now measurable with substantially improved precision. Most notably, the precision of $\dot{\omega}$ has improved by almost an order of magnitude, reflecting both the longer baseline and the high timing quality of the MeerKAT data. The Shapiro delay parameters $h_3$ and $\varsigma$ and the orbital period derivative $\dot{P}_\mathrm{b}$ have each improved by approximately a factor of two. These improvements have tightened the constraints on the component masses significantly. From the DDGR timing model, we obtain a pulsar mass of $m_\mathrm{p} = 1.3384(2)\,\mathrm{M}_\odot$, a companion mass of $m_\mathrm{c} = 1.3945(2)\,\mathrm{M}_\odot$, and a total system mass of $M_\mathrm{tot} = 2.7328884(36)\,\mathrm{M}_\odot$, all with significantly reduced uncertainties compared to prior work.

The most significant new result of this work is the detection of the relativistic angular deformation parameter $\delta_\theta$ for PSR~J1757$-$1854. PSR~J1757$-$1854 is now only the third DNS system in which this parameter has been detected, following the Hulse-Taylor binary PSR~B1913+16 \citep{Weisberg_Huang2016} and the double pulsar PSR~J0737$-$3039A/B \citep{kramer2021}. Strikingly, while the Hulse-Taylor pulsar required $\sim 40$\,yr of timing data to achieve this measurement \citep{Weisberg_Huang2016}, and PSR~J0737$-$3039A required $\sim 18$\,yr \citep{kramer2021}, PSR~J1757$-$1854 has yielded a detection in just 9 yrs, a testament to the extraordinary relativistic nature of the system and the transformative sensitivity of MeerKAT. We obtain a detection at $\sim 2\sigma$ significance when $\delta_\theta$ is sampled simultaneously with the Einstein delay $\gamma$, rising to $\sim 7\sigma$ when $\gamma$ is held fixed at its jointly sampled value. A critical technical point in this measurement is the strong correlation between $\delta_\theta$ and $\gamma$. As demonstrated previously in the context of the double pulsar \citep{kramer2021}, failing to account for this correlation leads to a biased estimate of $\gamma$, which in turn manifests as an apparent inconsistency in the mass-mass diagram. Indeed, we showed that the deviation of the $\gamma$ constraint from the common intersection of the other PK parameters in the initial DDH timing solution, a feature also noted by \cite{cameron2023new}, is naturally and fully explained by the unmodeled contribution of $\delta_\theta$. Once $\delta_\theta$ and $\gamma$ are sampled simultaneously, the mass-mass diagram becomes fully self-consistent, with all PK parameters intersecting in a common region. This underscores the importance of including $\delta_\theta$ in timing models for highly relativistic systems, where its omission can introduce systematic biases into the inferred masses and other PK parameters.

Even more interestingly, and despite its low relative precision, the measured value of $\delta_\theta$ has provided a powerful new means of constraining the spin-orbit geometry of PSR~J1757$-$1854. The observed value of $\delta_\theta$ contains a contribution from aberration that depends on the orientation of the pulsar spin axis relative to the orbital plane, parameterised by the precessing polar angles $\eta(t)$ and $\lambda(t)$. By computing the intrinsic relativistic contribution $\delta_\theta^\mathrm{intrinsic}$ for each of the four geometric solutions identified by \cite{cameron2023new} and comparing with the GR-predicted value $\delta_\theta^\mathrm{GR}$ derived from the component masses, we have been able to rule out two of the four solutions. Specifically, Solutions~1 and~3 yield values of $\delta_\theta^\mathrm{intrinsic}$ that are significantly inconsistent with $\delta_\theta^\mathrm{GR}$, while Solutions~2 and~4 remain consistent with the GR expectation. This conclusion is independently corroborated by the mass-mass diagrams constructed under each solution: the $\delta_\theta$ constraint band passes through the region of mutual consistency with all other PK parameters only for Solutions~2 and~4. These two solutions are additionally favoured by the evolutionary model of the system. This result demonstrates that measurements of $\delta_\theta$ provide a novel and powerful geometric discriminator between otherwise degenerate spin-orbit orientation solutions, complementing the constraints previously obtained from pulse profile evolution studies by \cite{cameron2023new}.

The observed orbital period derivative $\dot{P}_\mathrm{b}$, after subtraction of the kinematic contributions from the Shklovskii effect and Galactic differential acceleration, remains consistent with the GR prediction for gravitational wave-driven orbital decay across a wide range of plausible distances to the system. Although the dominant source of uncertainty in this test remains the poorly constrained distance to PSR~J1757$-$1854, currently estimated only from Galactic free electron density models, the excess period derivative $\dot{P}_\mathrm{b,exs}$ is consistent with zero. This result confirms that PSR~J1757$-$1854 continues to provide a valid, if distance-limited, test of the GR prediction for gravitational wave energy loss.

The exceptional precision achieved in the measurement of $\dot{\omega}$ in this work has made it necessary to assess contributions to the periastron advance beyond the standard leading-order 1PN approximation. We have evaluated the 2PN correction and the LT contribution arising from spin-orbit coupling between the pulsar's rotation and the orbital motion, which importantly already benefit from the geometric constraints imposed by our detection of $\delta_\theta$. These effects already impact estimates of the total mass with high significance. Correctly accounting for these higher-order terms will be essential for future high-precision tests of gravity with this system.

\subsection{Prospects for further investigation}

PSR~J1757$-$1854 remains one of the most scientifically productive and promising DNS systems known, and the results presented here open several important and interconnected avenues for future work.

\begin{itemize}

\item The continued monitoring of this system will substantially improve the significance of $\delta_\theta$ detection in the near future, in addition to the precision of $\gamma$. Furthermore, as the system precesses, the periastron will keep approaching superior conjunction. As shown by \cite{Weisberg_Huang2016}, this will greatly improve the precision of the Shapiro delay parameters. All this will greatly improve all GR tests with this system.

\item The geodetic precession of the pulsar spin axis, already established through pulse profile evolution by \cite{cameron2023new}, will continue to modify the pulse morphology over time. Continued monitoring will be important for improving the precision on the orientation of the spin axis of the pulsar. This effort will be further aided by the improved measurements of $\delta_\theta$.

\item Finally, an independent distance measurement to PSR~J1757$-$1854 remains a high priority, as it would reduce the uncertainty of the kinematic contributions to all secular PK parameters, thus improving the precision of the $\dot{P}_\mathrm{b}$ test.

\item A precise, independent measurement of the total system mass from $\dot{P}_\mathrm{b}$ could enable a future measurement of the LT contribution to $\dot{\omega}$, however, limitations on our knowledge of the Galactic potential make this unlikely for the foreseeable future, even if a very precise measurement of the distance became available.
The projected detection of $\dot{x}$ at $3\sigma$ significance no earlier than 2031 and of $\dot{e}$ no earlier than 2040 \citep{cameron2023new} remain important long-term goals. Again, knowing the orientation of the spin axis of the pulsar will be important for extracting the value of $\dot{x}$, as this will be correlated with a changing value of the aberration. The error in $\dot{x}$ in the present data is already better than the amplitude of the LT effect so in the next few years the error in $\dot{x}$ will decrease and LT contribution will grow rapidly.

\item Measuring either contribution of the Lense-Thirring effect
would provide direct observational access to the moment of inertia of the pulsar, $I_\mathrm{p}$, which is a sensitive probe of the neutron star equation of state and would represent only the second such measurement after the double pulsar \citep{kramer2021}. However, in order to extract the moment of inertia of the pulsar from the LT effect, knowing the orientation of the spin axis of the pulsar will be of paramount importance.

\item Knowledge of the orientation of the pulsar spin axis is also important for constraining the formation of PSR~J1757$-$1854. The angle between the pulsar spin axis and the orbital angular momentum vector arises from the natal kick imparted during the second supernova (SN). According to standard binary evolution theory (e.g. \citealt{Tauris_Heuvel2023}), the spin axis of the first-born NS (i.e. the observed, mildly recycled radio pulsar) is expected to be aligned with the orbital angular momentum axis following the accretion phase prior to the core collapse of the companion star. Any out-of-plane component of the natal kick in the second SN will tilt the orbital plane, thereby introducing a misalignment between the pulsar spin axis and the post-SN orbital angular momentum vector. Therefore, the observed spin orientation provides an important diagnostic of the properties of the second SN.
\end{itemize}

In summary, PSR~J1757$-$1854 stands as one of the most extreme and scientifically rich DNS binary systems accessible to current radio telescopes. The results presented here represent a substantial step forward in our understanding of this system and in our ability to use it as a laboratory for strong-field gravity tests. A detailed kinematic analysis of PSR~J1757$-$1854, including the system's velocity distribution and its implications for the binary's evolutionary history, will be presented in a future paper. This will also include polarimetric studies of the pulse profiles to constrain geodetic precession and infer the misalignment angle between the pulsar and orbital spins.


\section*{Acknowledgements}

The MeerKAT telescope is operated by the South African Radio Astronomy Observatory (SARAO), which is a facility of the National Research Foundation, an agency of the Department of Science and Innovation. SARAO acknowledges the ongoing advice and calibration of GPS systems by the National Metrology Institute of South Africa (NMISA) and the time space reference systems department of the Paris Observatory. PTUSE was developed with support from the Australian SKA Oﬃce and Swinburne University of Technology. This work made use of the OzSTAR national HPC facility at Swinburne University of Technology and the Hercules computing cluster of the Max Planck Institute for Radio Astronomy. MeerTime data are housed on the OzSTAR supercomputer. The OzSTAR program receives funding in part from the Astronomy National Collaborative Research Infrastructure Strategy (NCRIS) allocation provided by the Australian Government. We acknowledge the use of the ilifu cloud computing facility – www.ilifu.ac.za, a partnership between the University of Cape Town, the University of the Western Cape, Stellenbosch University, Sol Plaatje University and the Cape Peninsula University of Technology. The ilifu facility is supported by contributions from the Inter-University Institute for Data Intensive Astronomy (IDIA – a partnership between the University of Cape Town, the University of Pretoria and the University of the Western Cape), the Computational Biology division at UCT and the Data Intensive Research Initiative of South Africa (DIRISA). J.S., M.G. and A.W. acknowledge the support from the University of Cape Town Vice Chancellor’s Future Leaders 2030 Awards programme and the South African Research Chairs Initiative of the Department of Science and Technology and the National Research Foundation. V.V.K. acknowledges financial support from the European Research Council (ERC) Starting Grant ``COMPACT'' (Grant Agreement No. 101078094). M.A.M and V.B. are members of the NANOGrav Physics Frontiers Center (NSF award number 2020265). A.Po. acknowledges resources from the INAF Large Grant 2022 “GCjewels” (P.I. Andrea Possenti) approved with the Presidential Decree 30/2022. This work was supported in part by the “Italian Ministry of Foreign Affairs and International Cooperation”, grant number ZA23GR03, under the project "RADIOMAP- Science and technology pathways to MeerKAT+: the Italian and South African synergy". Parts of this research was conducted by the Australian Research Council Centre of Excellence for Gravitational Wave Discovery (OzGrav), through project number CE23010004 CE230100016. We thank Adam Coelen for performing independent checks on the analysis presented in this work, which helped confirm the robustness of our results. We acknowledge the contributions of Duncan Lorimer in observations with the GBT. We also acknowledge the useful comments from Thomas Tauris on the paper.


\section*{Data Availability}
 
This paper partly uses the publicly availble data from \cite{cameron2023new}, which can be found on the gSTAR Data Management and Collaboration Platform (gDMCP) at \url{http://dx.doi.org/10.26185/647e84c9b6a9b}. Part of the additional data is available from the Swinburne pulsar portal: \url{https://pulsars.org.au}. Any other data used here will be provided on reasonable request.



\bibliographystyle{mnras}
\bibliography{references} 



\appendix
\section{White noise model parameters}
Table~\ref{tab:noise_params} lists the white noise parameters: EFAC and EQUAD, determined from the Bayesian noise analysis using \textsc{temponest} for each telescope, receiver, and backend combination in the combined dataset. The EQUAD values are expressed in $\log_{10}$ units.
\begin{table}
\caption{EFAC and EQUAD noise parameters for each telescope/receiver/backend combination.}
\label{tab:noise_params}
\begin{center}
\begin{tabular}{lllcc}
\hline
\hline
Telescope & Receiver & Backend & EFAC & EQUAD \\
 & & & & (log$_{10}$) \\
\hline
\textit{Parkes} & MB20 & BPSR   & 0.8027 & $-8.9644$ \\
                &      & DFB4   & 1.1574 & $-6.7859$ \\
                &      & CASPSR & 1.0737 & $-9.4943$ \\
                & H-OH & DFB4 (fold)   & 4.8367 & $-4.6016$ \\
                &      & DFB4 (search) & 1.1296 & $-7.7312$ \\
                &      & CASPSR & 1.0272 & $-6.5204$ \\
                & UWL  & DFB4 (fold)   & 0.7255 & $-7.5152$ \\
                &      & DFB4 (search) & 0.9900 & $-8.4321$ \\
                &      & CASPSR & 0.6788 & $-7.1136$ \\
\hline
\textit{GBT}    & PF1-800 & GUPPI         & 1.0991 & $-4.9031$ \\
                & L-Band  & GUPPI         & 1.1414 & $-8.1120$ \\
                &         & VEGAS         & 1.0284 & $-5.2050$ \\
                & S-Band  & GUPPI         & 0.9719 & $-4.9777$ \\
                &         & GUPPI (INCOH) & 1.1602 & $-8.3727$ \\
                &         & VEGAS         & 0.9662 & $-5.0230$ \\
\hline
\textit{\textbf{MeerKAT}} & \textbf{L-Band} & \textbf{PTUSE} & \textbf{2.2446} & $\mathbf{-7.2721}$ \\
                          & \textbf{S-Band} & \textbf{PTUSE} & \textbf{0.8453} & $\mathbf{-4.8441}$ \\
\hline
\textit{\textbf{GBT -- New}} & \textbf{L-Band} & \textbf{VEGAS} & \textbf{0.9396} & $\mathbf{-4.5049}$ \\
\hline
\hline
\end{tabular}
\end{center}
\end{table}


\section{Moment of inertia of the pulsar}

To calculate the Lense-Thirring contribution to the periastron advance caused by the rotation of PSR~J1757$-$1854, one needs to know the spin of the pulsar and consequently its moment of inertia (MOI), $I_\mathrm{p}$. Given the precise measurement of the pulsar mass $m_\mathrm{p} = 1.3384(2)~\mathrm{M}_\odot$ obtained from our timing analysis (see Table~\ref{tab:J1757timingparameters}), the MOI is determined with high precision once an equation of state (EOS) has been chosen\footnote{In future, an observational determination of the MOI from PK parameters would, when combined with the precisely known pulsar mass, provide a constraint on the NS radius, and thereby on the EOS, rather than requiring an assumed EOS to determine the MOI.} to describe the pressure-density relation for neutron star matter. However, our knowledge of the properties of matter at supranuclear densities remains limited, leading to considerable uncertainty in the true EOS of neutron stars. This translates directly into a corresponding uncertainty in the MOI of a neutron star of given mass. 

The purpose of this section is to derive a simple probability distribution for $I_\mathrm{p}$ that (at least partially) reflects this imperfect knowledge of the internal density structure of PSR~J1757$-$1854. The ideal approach would be to follow a framework analogous to that of \citet{Koehn_2024}: construct a set of $10^5$ EOS candidates, calculate the MOI of PSR~J1757$-$1854 for each candidate, and apply nuclear and astrophysical constraints. This, however, would be far beyond the scope of the present work, given that in this paper we only need a rough estimate of the MOI and its uncertainty.

To considerably simplify the calculations, we build on the constraints on $R_{1.4}$ published by \citet{Koehn_2024}. $R_{14}$ is the radius of a canonical non-rotating 1.4\,$\mathrm{M}_\odot$ neutron star. As it depends on the EOS, the probability distribution for $R_{1.4}$ of \citet{Koehn_2024} can serve as a proxy for the probability of a given EOS.\footnote{To extract the probability distribution for $R_{1.4}$ from \citep{Koehn_2024}, we made use of the WebPlotDigitizer (URL: https://automeris.io).}
This in turn, can then be used to obtain a probability distribution for any other quantity calculated for a neutron star with that specific EOS (for any given mass supported by this EOS). For us this is the MOI $I_\mathrm{p}$ of PSR~J1757$-$1854, i.e.\ a neutron star with $1.338~\mathrm{M}_\odot$.

\begin{figure}
  \centering
  \includegraphics[scale=.325, trim={0 0 0 0},clip]{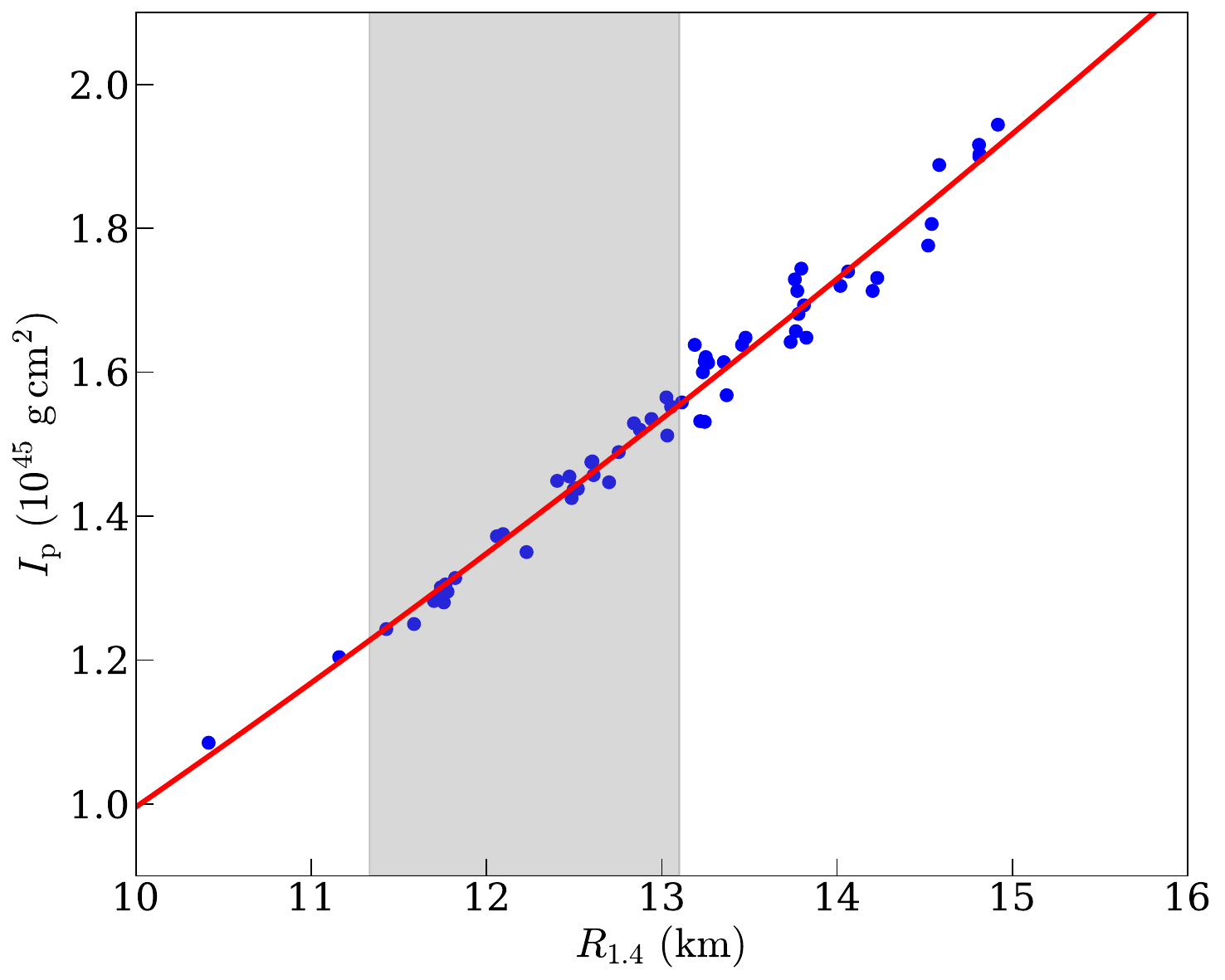}
  \caption{Moment of inertia of PSR~J1757$-$1854, computed for 62 different EOSs that all support a neutron star with at least $1.96~\mathrm{M}_\odot$\protect\footnotemark[\value{footnote}], shown as a function of $R_{1.4}$. $R_{1.4}$ is the calculated radius of a non-rotating neutron star of $1.4~\mathrm{M}_\odot$ and is used as parametrize the set of EOSs and, most importantly, place constraints on the range of EOSs (see text for details). For all these 62 EOSs we use the piecewise polytropic approach of \protect\cite{Read_2009} with parameters from Table~III of \protect\cite{Read_2009} and Table~V of \protect\cite{Kumar_2019}. The grey range indicates the conservative 95\% credibility interval of $R_{1.4}$ given by \protect\cite{Koehn_2024}. The red curve shows a quadratic least-squares-fit to the data points: $I_\mathrm{p}[10^{45}~\mathrm{g}~\mathrm{cm}^2] = 0.003638 \,(R_{1.4}[\mathrm{km}])^2 + 0.09612 \, (R_{1.4}[\mathrm{km}]) - 0.3292$.}
  \label{fig:MOI}
\end{figure}

To infer the probability distribution of $I_\mathrm{p}$, we employed a Monte Carlo sampling approach. For 62 EOSs, we computed pairs $(R_{1.4,i}, I_{\mathrm{p},i})$, we fitted a quadratic polynomial $\mathcal{P}_2(R_{1.4})$, to these pairs using least squares, which provided a satisfactory description of the EOS sample (see Fig~\ref{fig:MOI}). A total of $10^7$ values of $R_{1.4}$ were drawn from the conservative probability density for $R_{1.4}$ of \citet{Koehn_2024} (see Set A of their Fig.~26). Each $R_{1.4}$ was then mapped to a corresponding $I_\mathrm{p}$ via $I^{(j)} = \mathcal{P}_2(R_{1.4}^{(j)})$. The resulting distribution can be approximated well by a Gaussian distribution with 
\begin{equation}
    I_\mathrm{p} = (1.394 \pm 0.088) \times 10^{45} ~ \mathrm{g~cm^2} \,.
    \label{eq:MOI}
\end{equation}
In the main text, we use this (Gaussian) distribution to estimate the uncertainty in our knowledge of the Lense-Thirring contribution to the advance of periastron ($\dot\omega_\mathrm{LT}$). To some extent, Eq.~(\ref{eq:MOI}) can only serve as a qualitative estimate of the uncertainty in the MOI, since the analysis presented here is limited to the class of neutron-star EOSs that can be well approximated by the piecewise polytropic approach of \cite{Read_2009}.


\stepcounter{footnote}
\footnotetext[\value{footnote}]{EOSs with a maximum mass below this threshold are excluded with 95\% confidence by the mass measurement of \cite{Fonseca_2021}. The somewhat more model-dependent neutron star mass for PSR~J0952$-$0607 found by \cite{Romani_2022} excludes maximum neutron-star masses below 1.96~$\mathrm{M}_\odot$ with about 99\% confidence.}


\bsp	
\label{lastpage}
\end{document}